\documentclass[twocolumn,showpacs,prd,aps,amsmath,amssymb,nofootinbib]{revtex4-1}
\usepackage{hyperref}

\usepackage{graphicx,psfrag}
\usepackage{color}
\usepackage{amsmath,amsfonts,amssymb,amsthm}
\usepackage{hyperref}
\usepackage{times}
\usepackage{scrextend}

\usepackage{natbib}
\usepackage{verbatim}
\usepackage{hyperref}
\usepackage{float}
\usepackage{wasysym}
\usepackage{amssymb,graphicx}
\usepackage{epsfig}
\usepackage{psfrag}
\usepackage{dsfont}
\usepackage{amsfonts}
\usepackage{mathrsfs}
\usepackage{multirow}
\usepackage{times}

\begin{document}
\title{Simulating the Magnetorotational Collapse of Supermassive
  Stars: Incorporating Gas Pressure Perturbations and Different
  Rotation Profiles}
\author{Lunan Sun$^1$}
\author{Milton Ruiz$^1$}
\author{Stuart L. Shapiro$^{1,2}$}
\affiliation{$^1$
  Department of Physics, University of Illinois at Urbana-Champaign,
  Urbana, IL 61801\\
  $^{2}$Department of Astronomy \& NCSA, University of Illinois at
  Urbana-Champaign, Urbana, IL 61801}

\begin{abstract}
  Collapsing supermassive stars (SMSs) with masses $M \gtrsim 10^{4-6}M_\odot$
  have long been speculated to be the seeds that can grow and become supermassive
  black holes (SMBHs). We previously performed general relativistic magnetohydrodynamic
  (GRMHD) simulations of marginally stable $\Gamma = 4/3$ polytropes uniformly rotating
  at the mass-shedding limit and endowed initially with a dynamically unimportant dipole
  magnetic field to model the direct collapse of SMSs. These configurations are supported
  entirely by thermal radiation pressure and reliably model SMSs with $M \gtrsim
  10^{6}M_\odot$. We found that around $90\%$ of the initial stellar mass forms a spinning
  black hole (BH) remnant surrounded by a  massive, hot, magnetized torus,
  which eventually launches a magnetically-driven jet. SMSs could be therefore sources of ultra-long
  gamma-ray bursts (ULGRBs). Here we perform GRMHD simulations of $\Gamma \gtrsim 4/3$, polytropes to
  account for the perturbative role of gas pressure in SMSs with $M \lesssim 10^{6}M_\odot$. We also
  consider different initial stellar rotation profiles. The stars are initially seeded with a dynamically
  weak dipole magnetic field that is either confined to the stellar interior or extended from its
  interior into the  stellar exterior. 	We calculate the gravitational wave burst signal for the different
  cases. We find that the mass of the black hole remnant is $90\%-99\%$ of the initial stellar mass,
  depending sharply on $\Gamma -4/3$ as well as on the initial stellar rotation profile. After
  $t\sim 250-550M\approx 1-2\times 10^3(M/10^6M_\odot)$s following the appearance of the BH
  horizon, an incipient jet is launched and it lasts for $\sim 10^4-10^5(M/10^6M_\odot)$s,
  consistent with the duration of long gamma-ray bursts. Our numerical results suggest that the Blandford-Znajek
  mechanism powers the incipient jet. They are also in rough agreement with our recently
  proposed universal model that estimates accretion rates and electromagnetic (Poynting) luminosities
  that characterize magnetized BH-disk remnant systems that launch a jet.  This model helps explain why the
  outgoing electromagnetic luminosities computed for vastly different BH-disk formation scenarios all
  reside within a narrow range ($\sim 10^{52 \pm 1} \rm erg s^{-1}$), roughly independent of $M$.
\end{abstract}
\maketitle
\bigskip

\section{Introduction}
\label{section:Introduction}
The discovery of quasars at high cosmological redshifts, e.g. J1342+0928 at redshift $z=7.54$
\cite{BanVenMaz17}, J1120+0641 at redshift $z=7.09$~\cite{MorWarVen11}, and SDSS J0100+2802
at redshift $z=6.33$ \cite{2015Natur.518..512W}, strongly supports the idea that supermassive
black holes (SMBHs) with masses $M \gtrsim 10^9M_\odot$ exist in the early universe. At the same time,
these observations raise questions about how SMBHs
could be formed in less than a billion years after the Big Bang, as well as about their growth
processes (see~\cite{2017A&G....58c3.22S} for a recent review). A possible scenario to explain
the origin of SMBHs is provided by the collapse of supermassive stars (SMSs)
with  masses  $\gtrsim 10^{4}M_{\odot}$ to black holes (BHs) following
their quasistationary cooling  and  contraction evolution epochs. These seed BHs,
at large redshifts ($z \sim 10-15$), could grow through accretion and mergers to become SMBHs
\cite{Ree84,BegVolRee06,Beg10}. An alternative scenario is the collapse of Population III (Pop III)
stars with $M \sim 100-500 M_{\odot}$ at $z \sim 20$ (e.g.~\cite{Ree84,MadRee01,Vol10, Kar13,
  CheHegWoo14}). For less massive Pop III stars ($140M_{\odot} \lesssim M \lesssim 260M_{\odot}$),
the electron-positron pair instability would cause rapid stellar contraction and oxygen and silicon
burning would produce sufficient energy to reverse the collapse and form pair-instability
supernovae~\cite{BonArnCar84, HegWoo02a}. However, it is believed that with $M > 260 M_{\odot}$,
nuclear burning is not powerful enough to overcome the implosion by the pair instability and the star
would collapse to a BH (e.g.~\cite{RakShaZin67,BonArnCar84,Woo86,HegWoo02a}). As pointed out in
e.g.~\cite{Alvarez:2008vw}, a $100M_\odot$ seed BH that accretes at the Eddington limit with
$\sim 10\%$ radiative efficiency can grow to $M_{BH}\gtrsim 10^9M_\odot$ by $z=6.4$, but only if the
onset of accretion is at $z>20$.

Idealized SMSs are objects supported dominantly by radiation pressure $P_r$, which can be
well described by a~$\Gamma= 4/3$ adiabatic index, or an $n = 3$ polytropic equation of
state~\cite{ShaTeu79,BauSha99,SaiThoSha02}. SMSs are likely to be highly spinning and turbulent
viscosity induced by magnetic fields would keep them in uniform rotation~\cite{Bisnovatyi67,
  Wag69,ZelNov71, Sha00}. The critical configuration of a SMS at the mass-shedding
limit along a quasistationary evolution sequence is set by the onset of a relativistic radial
instability. It has been pointed out that the ratio of rotational kinetic energy and gravitational
potential energy $T/|W|$, the compaction parameters $R_p/M$, where $R_p$ is the polar radius, and
the dimensionless spin $J/M^2$ for this critical configuration are all independent of the initial
mass~\cite{BauSha99}. Such universality also applies to the BH-disk parameters after collapse, as
shown by analytic models and full general relativistic (GR) hydrodynamic simulations of marginally
unstable, uniformly rotating SMSs spinning at the mass-shedding limit~\cite{ShiSha02,ShaShi02,
  LiuShaSte07,SunPasRui17}. These have shown that the SMS remnant is a black hole
surrounded by a massive, hot accretion torus. The remnant black hole has a mass $M_{BH}$ of
about~$\sim 90\%$ of the initial stellar mass $M$ and spin $a_{BH}/M_{BH}\sim 0.70-0.75$.  GRMHD
simulations in which the SMS is threaded initially by a dynamically weak dipole magnetic field,
either confined or not to the stellar interior, have shown that the above parameters remain basically
unchanged. In the magnetized case, however, following the gravitational wave (GW) burst at collapse,
the BH--accretion disk remnant gives rise to a magnetically confined jet with an outgoing
electromagnetic (Poynting) luminosity $L_{EM}~\sim 10^{52\pm1}\rm erg/s$, consistent with typical
GRB luminosities~\cite{LiuShaSte07,SunPasRui17}. This feature may explain the recent detection
of high redshift ($z \sim 5.3-8.0$) GRBs reported from the Burst Alert Telescope (BAT) on
\textit{Swift}. It may indicate that some metal--free Pop III stars could also be the engines that
power long GRBs (see e.g.~\cite{swiftGRB140304A,swiftGRB090423}), as they are at the epoch when
Pop III stars reached formation peak~(see e.g.~\cite{TorFerSch07,JohVecKho13}). The jets also exhibit
universal characteristics independent of mass. We explained this universality~\cite{Sha17} by an
analytic model that estimates several key global parameters characterizing a BH--accretion disk
remnant that launches a magnetically--driven jet consistent with the Blandford-Znajek (BZ)
mechanism~\cite{BlaZna77}. The same universal model accounts for BH--disk systems formed either
through compact binary mergers (i.e. neutron star or black hole--neutron star binary mergers,
such as in~\cite{PasRuiSha15,RuiLanPas16}, or massive star collapse as in~\cite{LiuShaSte07,
  SunPasRui17}.)

Some numerical simulations have shown that the gravitational collapse could be overcome by thermonuclear
energy if the SMSs have non-zero metallicity. In~\cite{FulWooWea86} a series of nonrotating SMSs with
different metallicity $Z$ were studied analytically and numerically, and microphysical processes
including electron-positron pairs, rapid proton capture and neutrinos loss were considered. They
found that hydrogen burning by the CNO cycle would trigger the explosion with a metallicity as low as
$Z = 5 \times 10^{-3}$ and release $2 \times 10^{56} - 10^{57}\rm{erg}$ of energy for stellar masses of
$10^5-10^6 M_{\odot}$.  It is also been found that the critical metallicity triggering the explosion
increases with stellar masses. A similar result was found by~\cite{MonJanMul12}, in which a
nonrotating SMS with mass of $\sim 5 \times 10^5 M_{\odot}$ would explode if the metallicity is greater
than $7 \times 10^{-3}$. Additionally, they discovered that the metallicity threshold is lowered to
$\sim 1 \times 10^{-3}$ if the stars are uniformly rotating. However, whether the massive stars could
contain the threshold metallicity is questionable, especially for the first generation of stars born in
metal-free regions. Although an $1D$ simulation of the evolution of Pop III SMSs by ~\cite{CheHegWoo14}
has shown that a $5 \times  10^4 M_{\odot} $ star could explode as a thermonuclear supernova powered
by helium burning, various approximations assumed and grid limitations may have hindered the accuracy
of the simulations.

Although numerical calculations obtained from strictly radiation-dominated $n = 3$ SMS models provide
promising observational suggestions, the approximation and simplification of the model may neither
accurately describe a realistic progenitor, nor sufficiently display some important physical
characteristics during the evolution. For example, SMSs also contain gas pressure $P_g \ll P_r$,
which becomes increasingly important as the mass of the star decreases. This importance
is reflected in the adiabatic index and polytropic index of the star. For a SMS with
$M \sim 10^5M_{\odot}$ the effective adiabatic index is $\Gamma = 1.339$ or $n = 2.95$ while for
$M \sim 10^4 M_{\odot}$ these parameters are $\Gamma = 1.345$ or $n = 2.9$. Both the critical
configuration at the onset of collapse and the final BH-disk system following collapse are
extremely sensitive functions of $\Gamma -4/3$ or $n - 3$, as we showed in ~\cite{Sha04}. Hence to
reliably track the onset of instability and the fate of an unstable SMS with mass $\lesssim 10^6
M_{\odot}$ it is necessary to simulate collapse from the critical configuration found for
$\Gamma > 4/3$. We also note that recent GR semi-analytic calculations and  hydrodynamic
simulations~\cite{ShiUchSek16,Shibata:2016vzw,ButLimBau18} suggest that the SMS in the nuclear burning
phase may be better described by a polytropic EOS in the range $2.95 \lesssim n \lesssim 3$.

As a uniformly rotating SMS contracts during its quasistationary cooling phase, its angular velocity
increases until reaching the maximally rotating (mass-shedding) limit. It will continue evolving along
a mass-shedding sequence~\cite{BauSha99,NewSha00a,Sha03ua,ButLimBau18}, as turbulent viscosity
arising from magnetic field instabilities likely maintain uniform rotation. Nevertheless, two
alternative situations might arise in principle. First, if the initial gaseous angular momentum is
not sufficient prior to contraction, then it is possible that SMSs do not spin-up sufficiently
to reach the mass-shedding limit when the radial-instability is triggered. Second, if magnetic
effects are greatly suppressed, then uniform rotation would not be sustained by turbulent processes
during the contraction phase and instead angular momentum would be conserved on each concentric
cylindrical shell~\cite{BodOst73, Tas78}. As a result, the SMSs would become differentially rotating,
even if uniformly rotating initially~\cite{NewSha00a}. Thus, simulating SMS collapse with the star
rotating differentially is also of interest. GR hydrodynamic simulations of collapsing differentially
rotating, radially unstable SMS models were performed first by ~\cite{Sai04}, who found the collapse
to be similar to that of a uniformly rotating star. A differentially rotating, $n=3$ polytrope with
a toroidal shape was studied in~\cite{ZinSteHaw05}. It was found that such an object is unstable to
nonaxisymmetric modes and fragmentation occurs. Recently, the evolution has been extended to
$\Gamma \gtrsim 4/3$, $n \lesssim 3$ SMS models where an initial $m =2$--sinusoidal density
perturbation triggered fragmentation that eventually formed a binary  BH surrounded by a cloud of
gas~\cite{ReiOttAbd13}. However, this simulation did not begin from an initially quasiequilibrium
state. GRMHD simulations that incorporate magnetic fields have yet to be performed for
this fragmentation scenario. 

The aim of the paper is twofold. First, we extend our previous GRMHD calculations~\cite{SunPasRui17}
of collapsing SMSs described by $\Gamma = 4/3$, $n=3$ polytropes to $\Gamma \gtrsim 4/3$, $n
\lesssim 3$ polytropes to treat lower mass models with gas pressure perturbations. We also consider
the evolution of SMS models with different initial stellar rotation profiles. Our simulations might be
useful for interpreting future coincident detections of GW bursts with electromagnetic (EM)
counterpart radiation (multimessenger observations).  Multimessenger signatures  from  the  direct
collapse of a SMS and the subsequent accretion epoch have not
been explored to a great extent. The future detection of GW signals by detectors such as
LISA~\cite{ArmAud16,BabHanHus08}, in coincident with GRBs at very high redshift, would
provide evidence for the direct--collapse massive-star model for the seeds SMBHs.
We also would like to verify the viability of the unified model presented in
\cite{Sha17}, which derives a direct relation between the EM signal
strength and the BH-accretion disk parameters.

We find that the mass of the black hole remnant is between $90\%$ and $99\%$ of the initial mass of
the SMS, depending sharply on $\Gamma-4/3$ or $n-3$ as well as on the initial rotation profile. The
latter
can affect the ram pressure produced by fall-back debris and the ultimate emergence of the jet. After
$t\sim 250-550M\approx 1-2\times 10^3(M/10^6M_\odot)$s following the appearance of the black hole
horizon, an incipient jet is launched in the magnetized cases considered, and it is expected
to last for $\sim 10^4-10^5(M/10^6M_\odot)$s, consistent with the duration of long gamma-ray bursts
\cite{Gendre2013ApJ,StrGenAtt13}. The outgoing electromagnetic Poynting luminosity driven by the jet is
$L_{EM} \sim 10^{51-53} \rm erg/s$.  As we pointed out in~\cite{SunPasRui17}, if $1\% - 10\%$ of this power
is converted into gamma rays, they can be detected potentially by \textit{Swift} and \text{Fermi}~\cite{Gehrels:2013xd}.
Our results also suggest that the BZ mechanism powers the incipient jet. We find that the estimates provided by
our unified model in~\cite{Sha17} are consistent with our numerical results within an order of magnitude. Finally,
we also diagnose the possibility of the quasi-periodic GW signature in the BH-disk system arising from the
Papaloizou--Pringle Instability (PPI)~\cite{PapPri84} as suggested in ~\cite{KiuShiMon11}. However, we find that
only the initial GW burst is appreciable and that no prominent signature of a PPI is found. 

The paper is organized as follows. In Sec. ~\ref{sec:ana_mod}, we summarize analytic calculations which model SMSs
with different characteristic masses and rotation profiles.  In Sec.~\ref{sec:method},  we discuss how the initial SMS
models are implemented numerically. We also describe the numerical methods used, as well as a number of diagnostic
quantities that we use to verify the reliability of our calculations.  In  Sec.~\ref{sec:results} we discuss
our results and compare them with our analytic model in~\cite{Sha17}. Finally, we summarize our conclusion and
propose future work in Sec.~\ref{sec:Summ-Conc}. Throughout the paper, we use geometrized units $c =G = 1$ unless
otherwise specified.
%

\section{Analytic Model}
\label{sec:ana_mod}
In this section we review key features of analytic SMS models described by polytropes with
different polytropic indices and rotation profiles. In Sec.~\ref{subsec:Mass}, we show how the
effective adiabatic (or polytropic) index of a SMS scales with mass when gas pressure perturbations
are included along with the dominant radiative pressure. In Sec.~\ref{subsec:Rotation}, we describe
the relation between angular velocity and the equatorial radius for uniformly rotating
stars, and we give the differential rotation profile used in one of our numerical models.

\subsection{Characteristic masses}
\label{subsec:Mass}
\begin{figure}[h]
\includegraphics[scale=0.1]{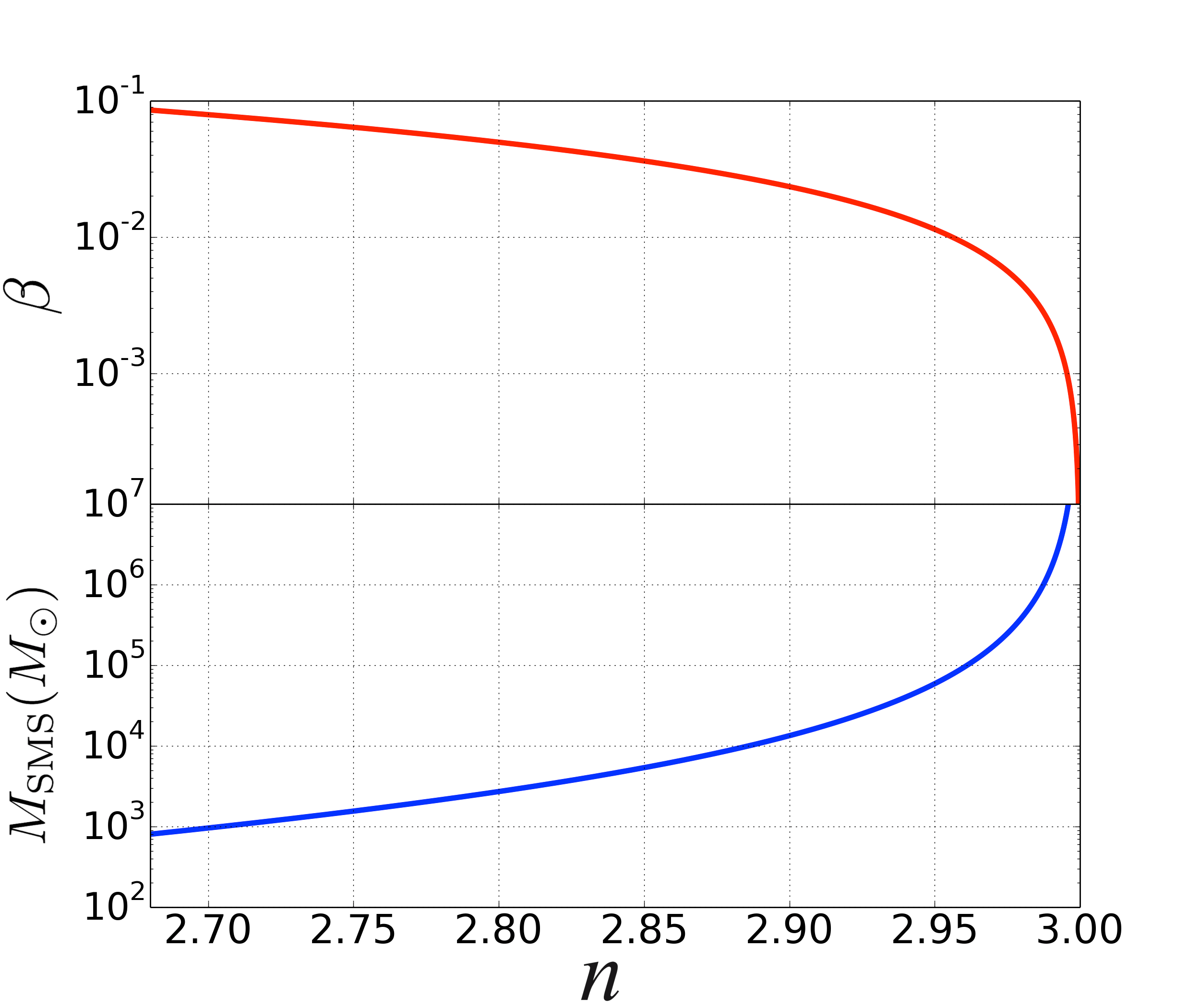}
\caption{\label{fig:M_vs_n} Gas-to-radiation pressure ratio $\beta$ (upper panel) and SMS mass (lower panel)
  as a function of polytropic index $n$ using Eqs.(\ref{eq:Gamma}) and (\ref{eq:nvsmass}). For $n$ within the range
  where $0 < \beta < 0.1$, gas pressure is a small perturbation, yet the mass varies by orders of magnitude.}
\end{figure}

Containing both radiation and gas pressure, a highly convective core maintains constant entropy
of the stellar interior~\cite{BonArnCar84, ShiUchSek16}.  Therefore, from the first law of
thermodynamics, a SMS can be modeled approximately by a polytrope with $P \propto \rho_0)^\Gamma$,
where $P$ is the pressure, and $\rho_0$ is the rest-mass density,
\begin{equation}
\label{eq:Gamma}
\Gamma = \frac{4}{3} + \frac{\beta(4+\beta)}{3(1+\beta)(8+\beta)} =  \frac{4}{3} + \frac{\beta}{6} + \mathcal{O} (\beta^2),
\end{equation}
and $\beta \equiv P_g /P_r$ is the ratio between the gas and the radiation pressure (see, e.g.,~\cite{Edd18a, Cha39,
  BonArnCar84, ShiUchSek16, ButLimBau18},  also see Problem 17.3 in ~\cite{ShaTeu83a} and Problem 2.26 in~\cite{Cla83}).
For radiation-dominated stars,  $\beta \ll 1$ is directly related to the radiation entropy $s_r$ and to the mass of a SMS.
To lowest order, and assuming stars consist of hydrogen only, we have
\begin{equation}
\beta= \frac{8 k_B}{s_r} = 8 .485 \left(\frac{M}{M_{\odot}}\right)^{-1/2},
\end{equation}
where $k_B$ is Boltzmann's constant. The relation between adiabatic index $\Gamma$ and $M$ to first order in $\beta$ is
\begin{equation}
\label{eq:Gammavsmass}
\Gamma -\frac{4}{3}\approx 1.414\left(\frac{M}{M_{\odot}}\right)^{-1/2},
\end{equation}
or, in terms of polytropic index $n\equiv 1/(\Gamma-1)$

\begin{equation}
\label{eq:nvsmass}
n \approx \frac{3}{1+4.242\left(\frac{M}{M_{\odot}}\right)^{-1/2}}.
\end{equation}
Fig.~\ref{fig:M_vs_n} displays $\beta$ and the mass of a SMS as a function of polytropic index
$n$ for $0<\beta < 0.1$. As $n$ decreases by a small amount, the resulting SMS mass drops by orders
of magnitude. A more detailed analysis of $\Gamma$ versus $M$ considering different components of
the plasma inside the star is proposed in ~\cite{ShiUchSek16}, which is consistent with the analysis above.
\subsection{Rotation profiles}
\label{subsec:Rotation}
Full discussion of an uniformly rotating, pressure dominated SMS is contained
in~\cite{BauSha99a} and~\cite{Sha04} in the Newtonian, Roche approximation.
There it is shown that the angular
frequency at the mass-shedding limit, where matter at the equator has no outward
support from pressure but is instead supported exclusively by centrifugal
forces and therefore follows a circular geodesic, satisfies
\begin{equation}
\label{eq:Omega_shedd}
\Omega_{\rm shedd} = \left(\frac{M}{R^3_{eq}}\right)^{1/2},
\end{equation}
where $R_{eq}$ is the equatorial radius.  
Integrating the hydrostatic equilibrium equation for a spherical stellar model in the Newtonian limit, we obtain (see Eq.~4 in~\cite{Sha04})
\begin{equation}
\label{eq:Omega_sheddII}
(n+1)\,\frac{P}{\rho}-\frac{M}{r}-\frac{1}{2}\,\Omega^2r^2\sin^2\theta=H\,.
\end{equation}
Here $H$ is a constant of integration. The angular velocity of an uniformly rotating star less than the mass-shedding limit can be described by $\Omega=\alpha\,\Omega_{\rm shedd}$, where $\alpha$ is a
spin-down factor that measures the deviation from the mass-shedding limit. Equating the values of Eq.~\ref{eq:Omega_sheddII} calculated at the pole and the equator, and assume that $R_{pol}$ of an uniformly rotating star is the
same as the nonrotating case~\footnote{This assumption was shown numerically to be very accurate, see e.g.~\cite{1973MNRAS.164....1P}}, 
we find that the ratio between
equatorial and polar radius of uniformly rotating Newtonian polytropes satisfies
\begin{equation}
\label{eq:alpha_rot}
  \frac{\alpha^2}{2}\left(\frac{2}{3}\right)^{3} \left(\frac{R_{eq}}{R_{pol}}\right)^3
  - \left(\frac{R_{eq}}{R_{pol}}\right) +1 = 0\,.
\end{equation}

We treated the collapse of a uniformly rotating, marginally unstable SMS with $n=3$ at mass-shedding
in~\cite{SunPasRui17}. For this case $\alpha = 1$, for which $R_{pol} = 2R_{eq}/3$. Here we consider the collapse of uniformly rotating, marginally unstable
configurations with $n=2.9$ and $n=2.95$ at the mass-shedding limit $\alpha = 1$. We also treat a $n = 2.9$ configuration at a smaller spin  $\alpha = 0.75$. We choose the smaller
$n$ in part to explore the effects of gas-pressure perturbations and in part to evolve a configuration
of smaller compaction and hence shorter dynamical and integration timescale. We use the approximate Newtonian model described above to provide input parameters for $R_{pol}/R_{eq}$ for insertion in our relativistic equilibrium code~\cite{CooShaTeu92,CooShaTeu94,Cook1994ApJ} to build a stable, uniformly rotating star. Our numerical solution is more accurate than the approximate Newtonian Roche model described by
Eq.~\ref{eq:alpha_rot}, although the discrepancy is not large even in the most compact case. For example, for $n = 2.9$ and $R_{pol}/R_{eq} = 0.89$, the numerically accurate GR value for $\alpha$ is 0.75, while Eq.~\ref{eq:alpha_rot} gives 0.77.

We also consider a differentially rotating configuration at the onset of instability. It is
defined by~\cite{DueLiuSha04, Sai04,ZinSteHaw06,ReiOttAbd13}
\begin{equation}
  u^tu_\phi=\frac{ R^2_{eq}}{9}\,(\Omega_c - \Omega)\,,
  \label{eq:rotation_pro}
\end{equation}
in the relativistic regime, where $\Omega = \Omega (\varpi)$ is the angular velocity
of the fluid, $\Omega_c$ is the angular velocity at the stellar center, and the $u^i$ are 4-velocity
components. In the Newtonian limit, Eq.~(\ref{eq:rotation_pro}) reduces to:
\begin{equation}
\Omega = \frac{\Omega_c}{ 1+ \frac{9\varpi^2}{ R^2_{eq}}}
\end{equation}
where $\varpi^2 = x^2+y^2$ is the distance from the rotation axis, with the center
of mass at the origin.
%
\begin{table*}[!ht]
  \caption{Summary of initial star parameters. Nondimensional quantities which have been rescaled
    with the polytropic gas constant $K$, are denoted with a bar. In all the magnetized stars the
    magnetic-to-rotational-kinetic-energy ratio is $0.1$.
    Columns show the polytropic index $n=1/(1-\Gamma)$, the characteristic mass $M_{\star}$ for which this
    index is most appropriate, the central rest-mass density $\bar{\rho}_{0,c}$,
    the ADM mass $\bar{M}_{ADM}$, the polar-to-equatorial radius ratio $R_{p}/R_{eq}$ , the
    equatorial radius $R_{eq}$, the dimensionless angular momentum $J/M_{ADM}^2$,
    the initial magnetic field configuration, the averaged magnetic field strength
    $\left<B\right>=\sqrt{8\,\pi\mathcal{M}/V_s}$, where $\mathcal{M}$ is the total magnetic energy and $V_s=\int \sqrt{\gamma}d^3x$
    is the initial proper volume of the star. 
      \label{tab:table1}}
\begin{ruledtabular}
\begin{tabular}{ccccccccccccc}
  Case                & $n$ &  $M_{\star}/M_{\odot}$& $\bar{\rho}_{0,c}$ \footnote{$\bar{\rho}_{0,c} = \rho_{0,c}K^n$, where $K = P/ \rho^{\Gamma}_0$, $\Gamma = 1+ \frac{1}{n}$ (for $K$ in cgs units, see~\cite{ShaTeu83a}, Eq. 17.2.6).}  & $\bar{M}_{ADM}$ \footnote{$\bar{M}_{ADM} = M_{ADM}K^{-n/2}$} & $R_{p}/R_{eq}$  & $R_{eq}/M_{ADM}$&    $J/M_{ADM}^2$  &   B-field & $\left<B\right>\times(M/10^6M_\odot)$ &\\

n3-HYD\footnote{Uniformly rotating star spinning at the mass-shedding limit.}     & 3  & $\gtrsim 10^6 $ & $7.7\times 10^{-9}$  & 4.57       & 0.67      &  625      & 0.96    &   None&  0       \\
n3-INT$^c$           & 3 & $\gtrsim 10^6 $  & $7.7\times 10^{-9}$  & 4.57      & 0.67     &  625     &     0.96    &    Int.        & $6.5\times 10^6$G                                   \\
n3-EXTINT$^c$        & 3 & $\gtrsim 10^6 $  & $7.7\times 10^{-9}$  & 4.57      & 0.67     &  625     &     0.96    &    Int .       & $6.5\times 10^6$G                                 \\
n295-EXTINT$^c$      &2.95 & $\sim 10^5 $ & $1.04\times 10^{-7}$ & 3.84      & 0.67     &  286     &     0.68    &    Int. + Ext. & $1.5\times 10^7$G                                \\
n29-HYD$^c$          &2.9  & $\sim 10^4 $ & $5.66\times 10^{-7}$ & 3.30      & 0.67     &  175     &     0.56    &    None        & 0                                                 \\
n29-INT$^c$          &2.9 & $\sim 10^4 $  & $5.66\times 10^{-7}$ & 3.30      & 0.67     &  175     &     0.56    &    Int.        & $4.7\times 10^7$G                                  \\
n29-EXTINT$^c$       &2.9  & $\sim 10^4 $ & $5.66\times 10^{-7}$ & 3.30      & 0.67     &  175     &     0.56    &    Int.+ Ext.  & $4.7\times 10^7$G                                  \\
n29-EXTINT-0.75SPIN\footnote{Uniformly rotating star spinning at 75\% of the mass-shedding limit.} &2.9  & $\sim 10^4 $ & $2.6\times 10^{-7}$& 3.26                  & 0.89      &  174      &  0.45  &  Int.+ Ext.& $2.7\times 10^7$G      \\
n29-EXTINT-DIFF\footnote{Differentially rotating star with the initial rotation profile
  given by Eq.~(\ref{eq:rotation_pro}).}&   2.9  & $\sim10^4 $ & $1.77\times 10^{-7}$ & 3.88           & 0.67      &   170     &   1.48   &     Int. + Ext.& $1.6\times 10^8$G     \\
     \hline
\end{tabular}
\end{ruledtabular}
\end{table*}

%
\section{Methods}
\label{sec:method}
In this section we begin with a summary of the numerical approach and code we employ for solving
GRMHD equations. A detailed description can be found in~\cite{Etienne:2010ui,EtVpas12}.
In~Sec.~\ref{subsec:ID} we describe our initial data. In particular, we discuss how we build our
initial SMS models, including the initial rotation profile and  the magnetic field
configuration seeded in the SMS. In~Sec.~\ref{subsec:grid} we review the resolution and grid
structure used during the different epochs of the stellar evolution. Finally, in Sec.
\ref{subsec:diag} we describe our standard tools to diagnose the numerical simulations.
%
\subsection{Numerical setup}
We use the moving-grid mesh refinement Illinois GRMHD code embedded in the {{\tt Cactus}\footnote{http://
    www.cactuscode.org}/{\tt Carpet}\footnote{http://www.carpetcode.org}} infrastructure. 
The code has been
extensively tested and used to study various scenarios, including magnetized compact object mergers and stellar
collapse, leading to magnetized accretion disks and in some cases the formation of jets~(see e.g.\cite{EtVpas12,Paschalidis:2013jsa,
  SunPasRui17,KhaPasRui18} and references  therein).

The Illinois GRMHD code evolves the spacetime metric by solving
Baumgarte--Shapiro--Shibata--Nakamura (BSSN) formulation of the Einstein's equations~\cite{ShiNak95, BauSha98b},
coupled to moving puncture gauge conditions~\cite{BakCenCho05,CamLouMar05} with the equation for
the shift vector in first-order form (see e.g.~\cite{HinBuoBoy13,RuiHilBer10}). Depending on the grid structure
and system properties for the different cases, the shift parameter $\eta$ is set between
$3.26/M$ and $3.89/M$, where $M$ is the ADM mass of the system. The code  solves the equations in a flux
conservative formulation [see Eqs.(27)-(29) in~\cite{Etienne:2010ui}]
via a high-resolution shock capturing method \cite{DueLiuSha05}. To guarantee that the magnetic field
remains divergenceless, the code solves the magnetic induction equation by introducing a vector potential
[see Eqs. (8)-(9) in \cite{EtVpas12}]. We adopt the generalized Lorenz gauge~\cite{EtVpas12,FarGolPas12}
to close Maxwell's equations. This gauge is chosen so that the development of spurious magnetic fields that
arise due to interpolations across AMR levels can be avoided; for details see~\cite{EtVpas12}.
The GRMHD evolution equations are evolved by employing a $\Gamma$-law
EOS, $P=(\Gamma-1)\,\epsilon\,\rho_0$, where $\Gamma\gtrsim 4/3$,  and $\epsilon$ and $\rho_0$ are
the specific internal energy and the rest-mass density, respectively.
%
\begin{figure*}[t]
\centering
\includegraphics[width=0.49\textwidth]{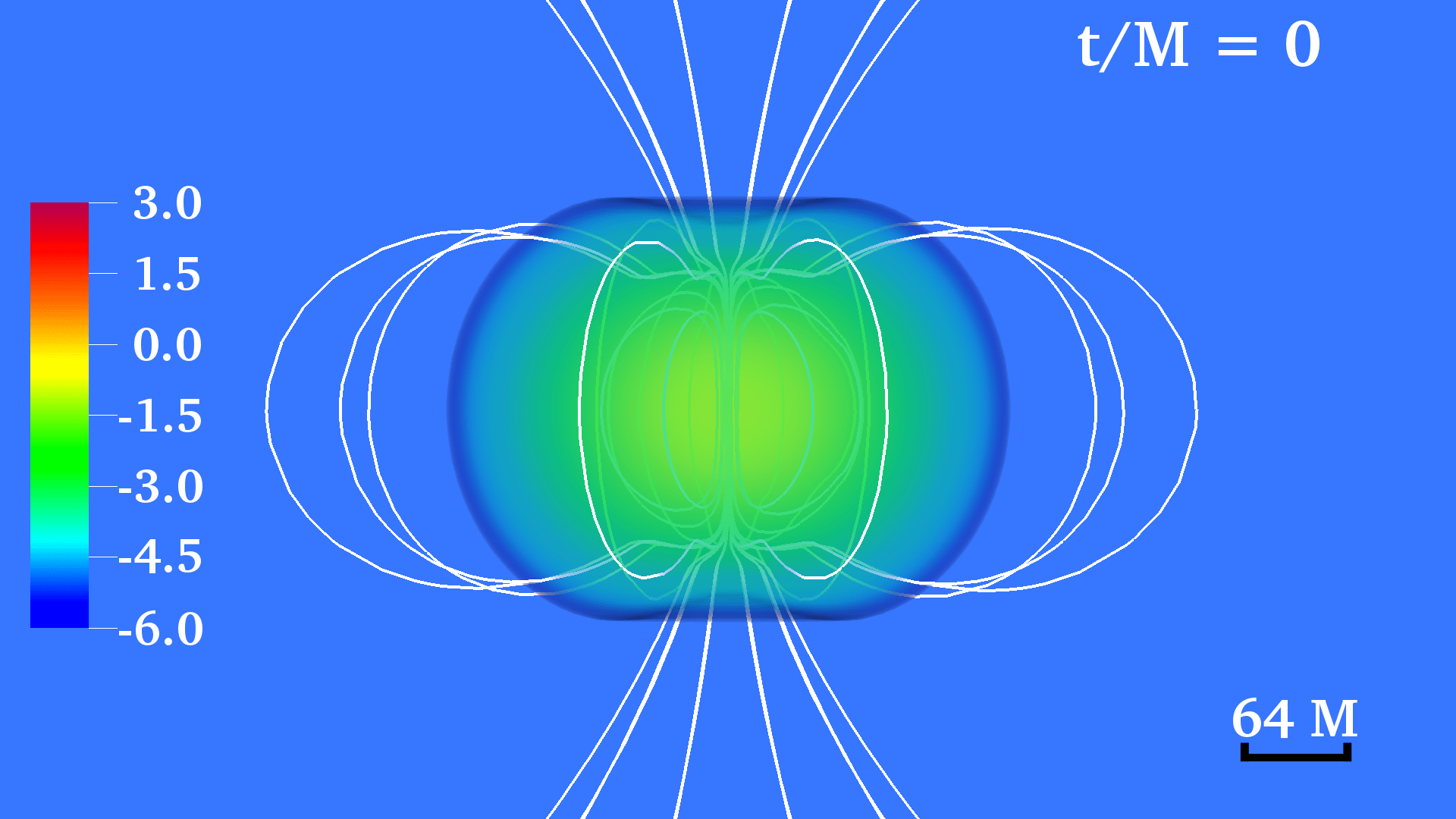}
\includegraphics[width=0.49\textwidth]{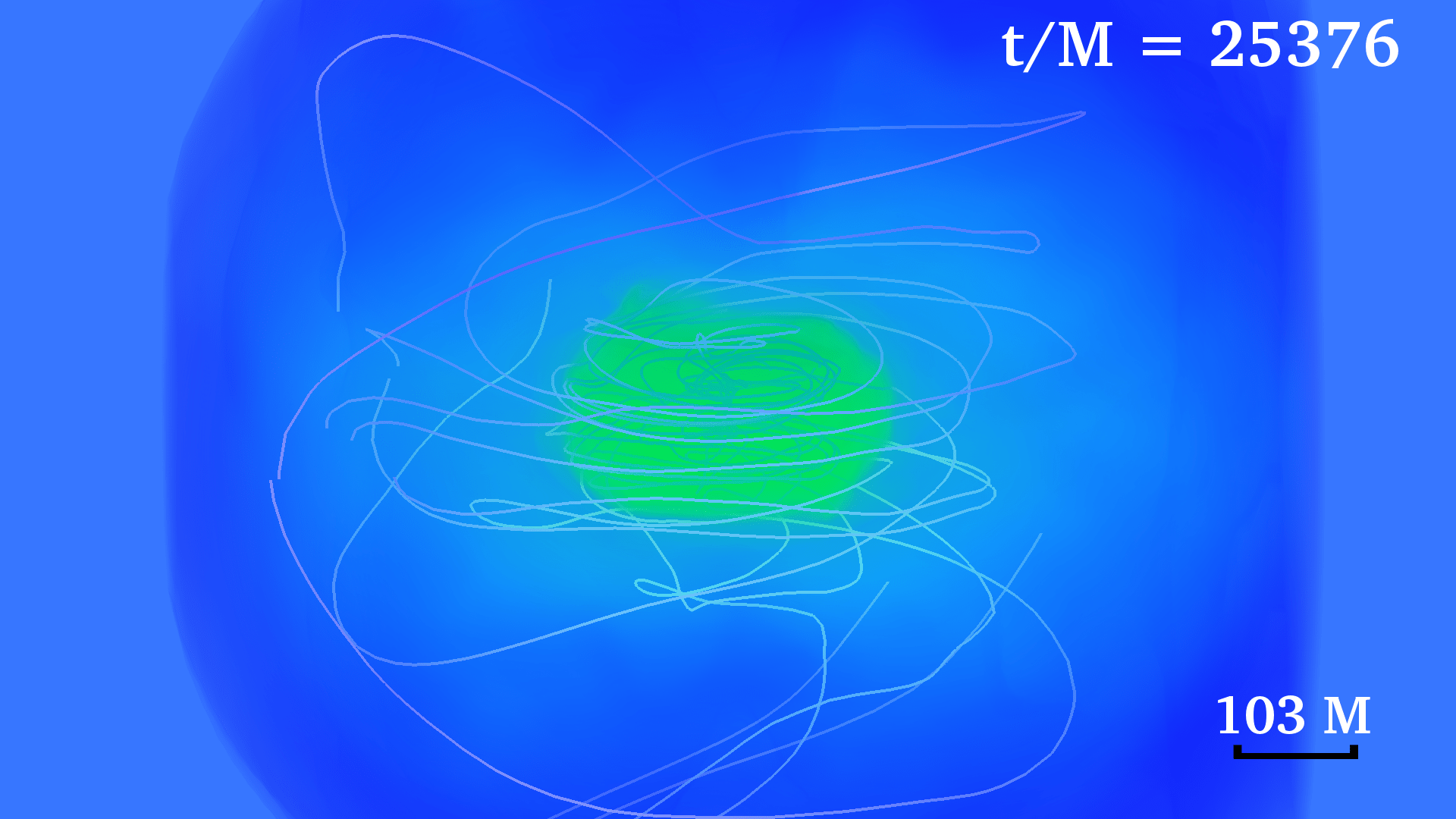}
\includegraphics[width=0.49\textwidth]{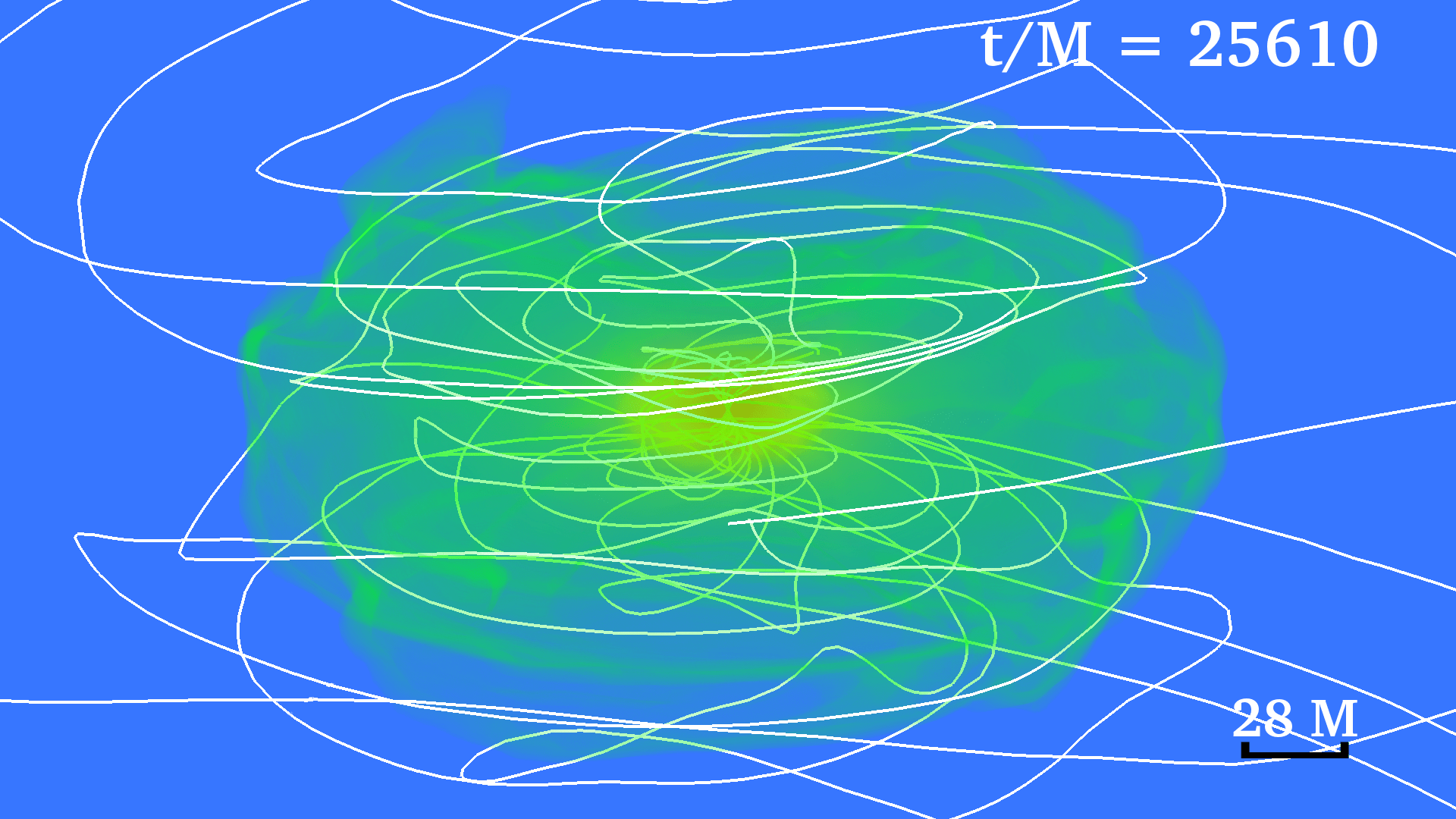}
\includegraphics[width=0.49\textwidth]{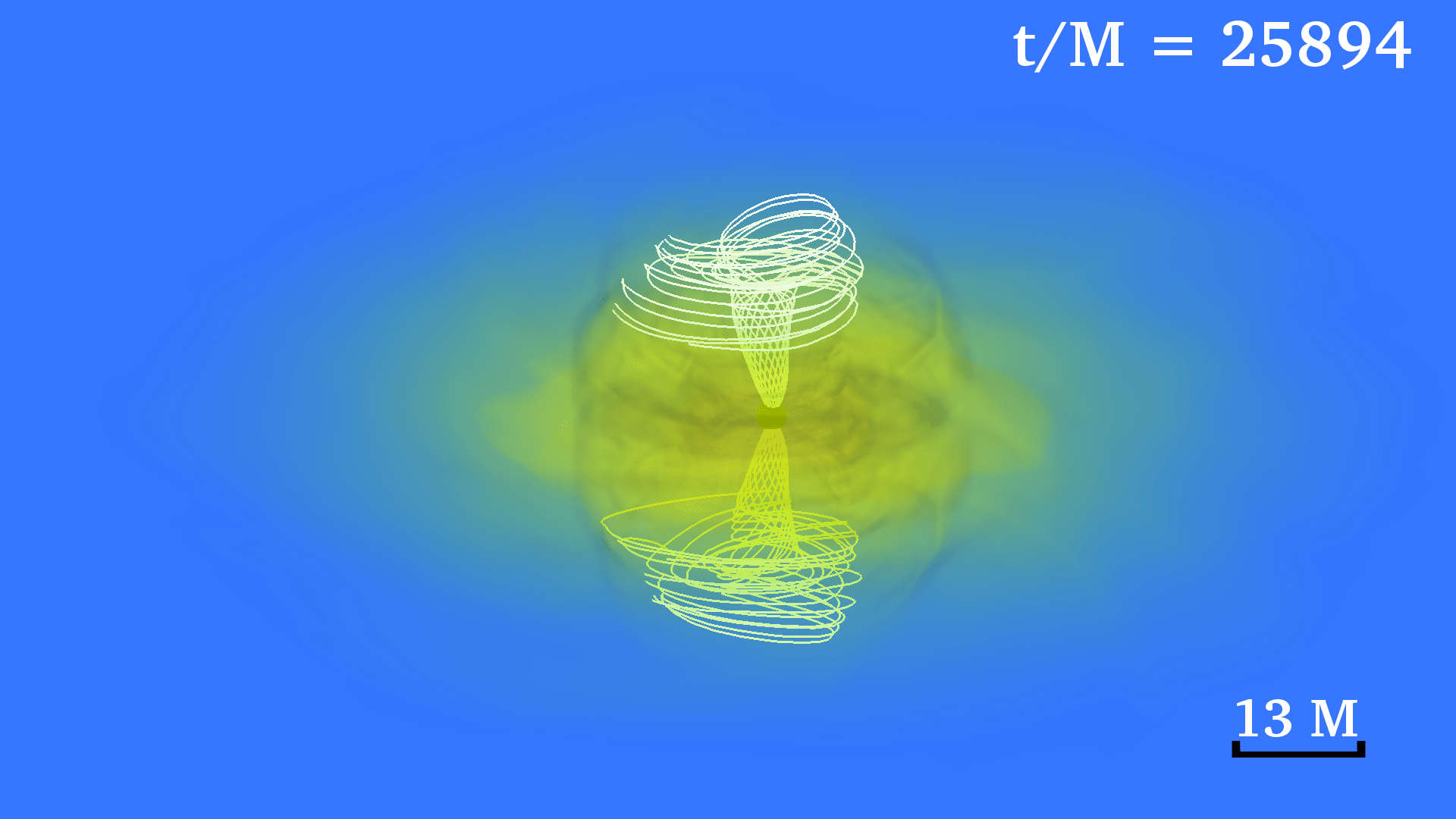}
\includegraphics[width=0.49\textwidth]{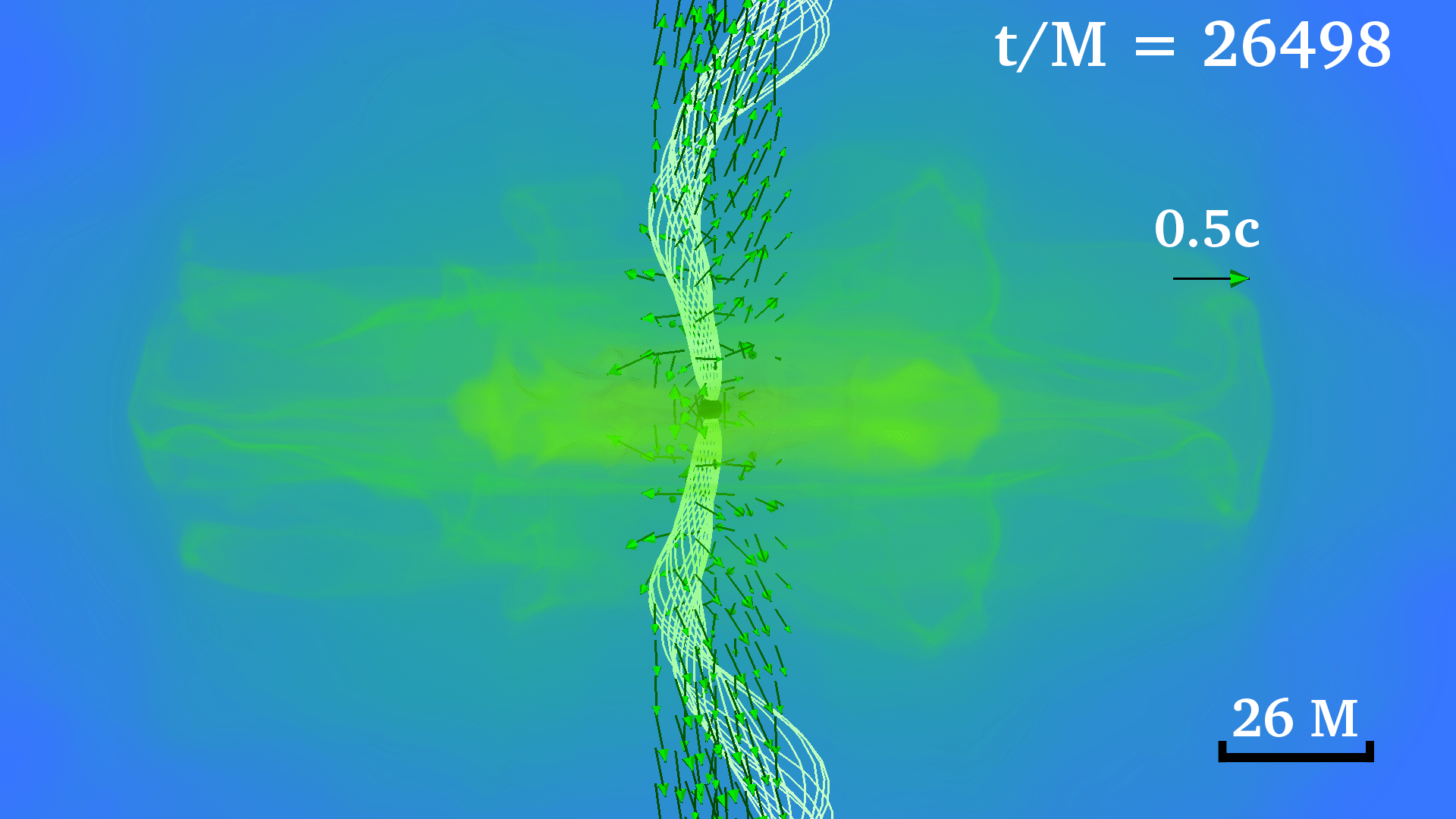}
\includegraphics[width=0.49\textwidth]{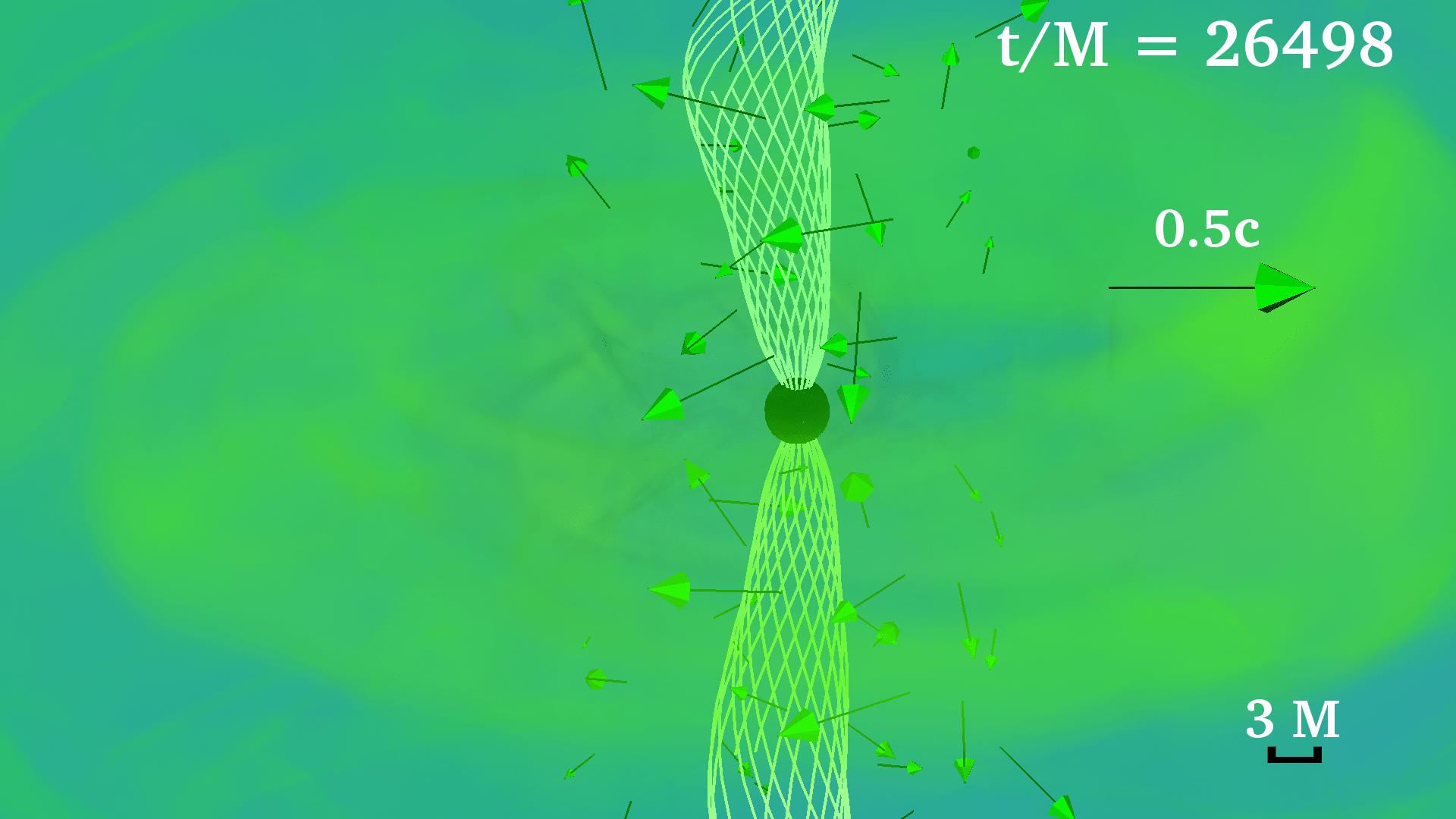}
\caption{\label{fig:Evolution_diff} 3D volume rendering of the rest-mass density normalized to its initial maximum
  value $\rho_{0,max}={1.66 (M/10^6M_{\odot})^{-2} \rm {g cm^{-3}}}$ at select times for the n29-EXTINT-DIFF case
  (see table~\ref{tab:table1}).  Solid lines indicate the magnetic field lines and arrows show plasma velocities
  with length proportional to their magnitude. The bottom left panel displays the collimated, helical magnetic
  field and outgoing plasma, whose zoomed-in view near the horizon is shown in the bottom right panel. Here
  $M=4.9 (M/10^6M_\odot){\rm s}=1.47\times 10^6(M/10^6M_\odot)$km.}
\end{figure*}
%
\subsection{Initial data}
\label{subsec:ID}
It is believed that SMSs form when colliding gas residing in metal--, dust--, and $H_2$--poor halos
build up sufficient radiation pressure to inhibit fragmentation and the formation of
small stars~\cite{Hai12,Regan14,Ress84,Gnedin:2001ey}. As thermal emission and turbulence driven by magnetic viscosity
take place, the star  shrinks and spins up to the  mass-shedding limit~\cite{Bisnovatyi67,ZelNov71,
  Shibata:2016vxy}. It then evolves in a quasistationary manner  until reaching the onset of relativistic radial
instability and eventually collapses to form a seed of a SMBH~\cite{BauSha99}. It also has been argued, that massive
stars with $M\gtrsim 10^2 M_\odot$and sufficiently low metallicity (Pop III stars) may be the progenitors of SMBHs, if
mass--loss mechanisms such as nuclear--powered radial pulsations and the electron-positron pair instability on the
main sequence are suppressed~\cite{HegWoo02,Baraffe:2000dp}.
Here, we consider SMSs described by a marginally unstable polytrope spinning at the mass-shedding limit characterized
by a polytopic index $n = 2.95$ and $n = 2.9$ (Table ~\ref{tab:table1}). Compared to $n=3$ polytropes which better
characterize SMSs with $M \gtrsim 10^6 M_{\odot}$, they correspond to SMSs with smaller characteristic mass of $10^5 M_{\odot}$
and $10^4 M_{\odot}$, respectively, according to Eq.~(\ref{eq:nvsmass}) and Fig.~\ref{fig:M_vs_n}. In order to study the
effects of the initial rotation profile, we model the uniformly rotating SMSs initial configuration at mass-shedding
and with 0.75 of the corresponding mass-shedding angular velocity. For the latter we set $\alpha = 0.75$, which gives $R_{pol}/R_{eq} \approx 0.89$. Finally we also consider differentially rotating stars
with an initial rotation profile given by Eq.~(\ref{eq:rotation_pro}).

To determine the central density $\rho_c$ of the marginally unstable stellar models spinning
at the mass-shedding limit for a given polytropic index $n$, we solve Eqs.$(17)$ and  $(18)$ along
with the constraint in Eq. $(19)$  in~\cite{Sha04}. Note that the configurations described by such a soft EOS ($n\approx 3$) are
low compaction stars.
Given the central density $\rho_c$ and the above polar-to-equatorial radius
ratio, we build the above rotating stellar configurations  with  the relativistic rotating star code described in
\cite{CooShaTeu92,CooShaTeu94,Cook1994ApJ}. 

\begin{figure*}[ht!]
\includegraphics[height=0.28\textwidth]{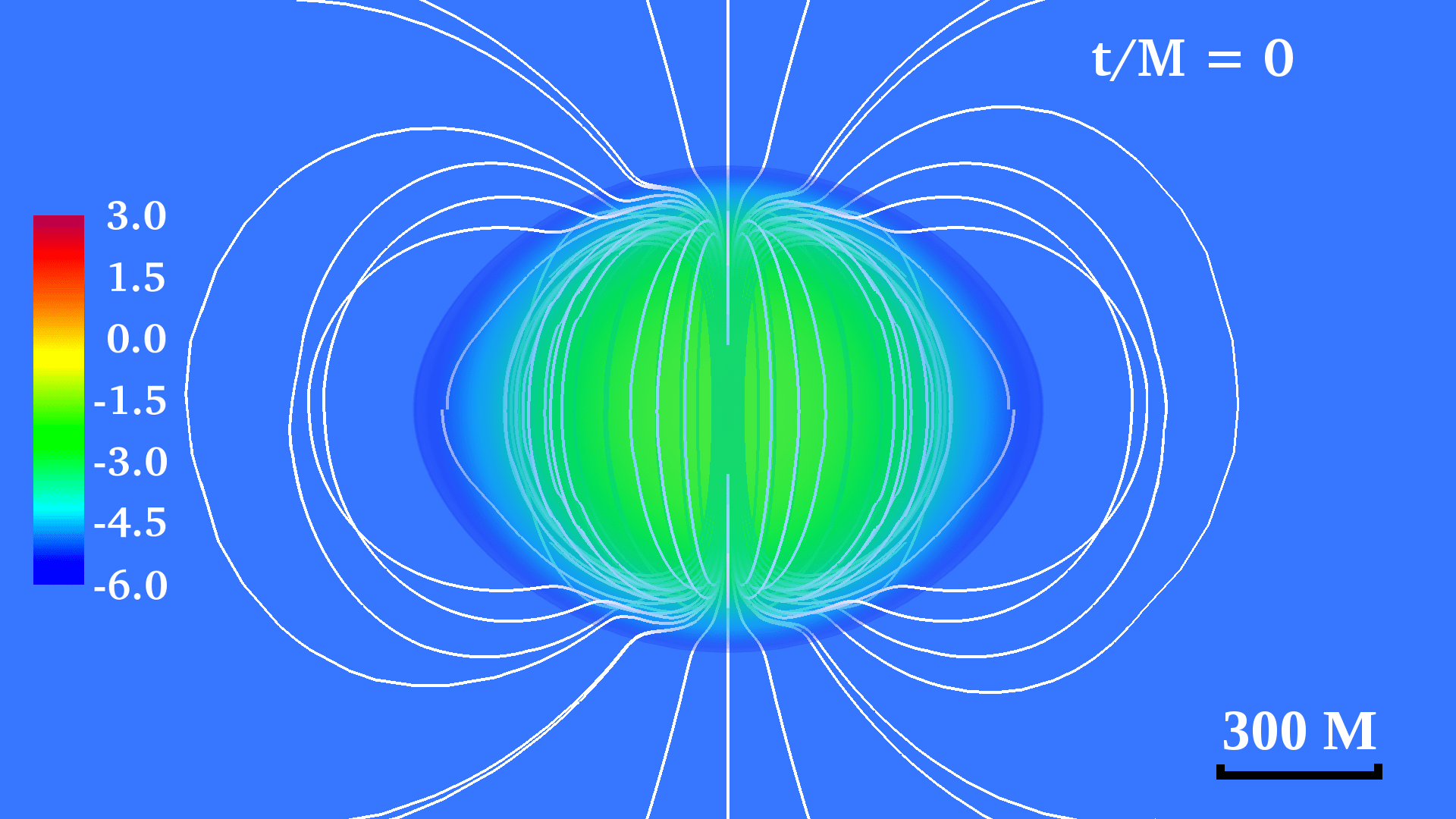}
\includegraphics[height=0.28\textwidth]{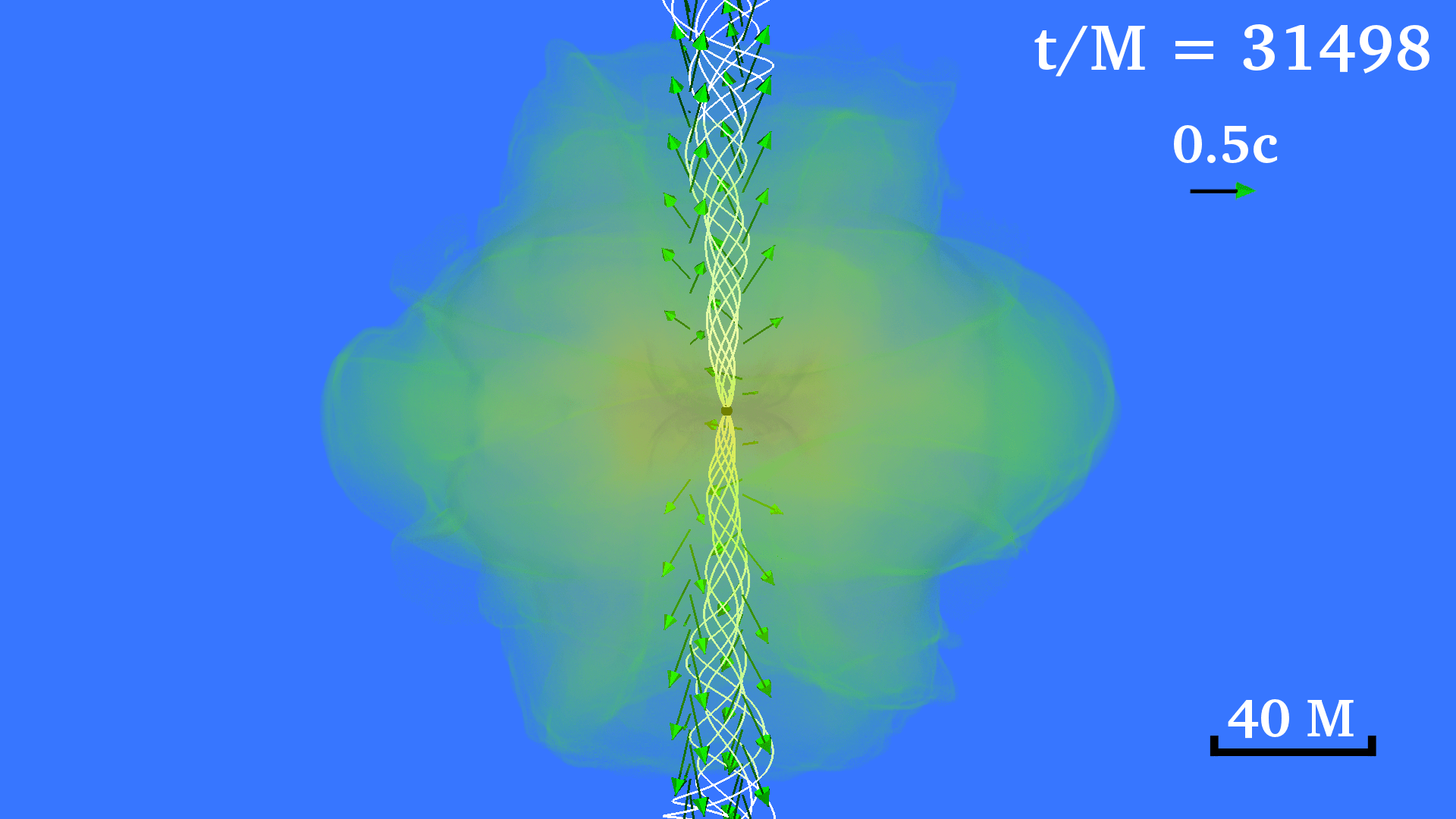}
\includegraphics[height=0.28\textwidth]{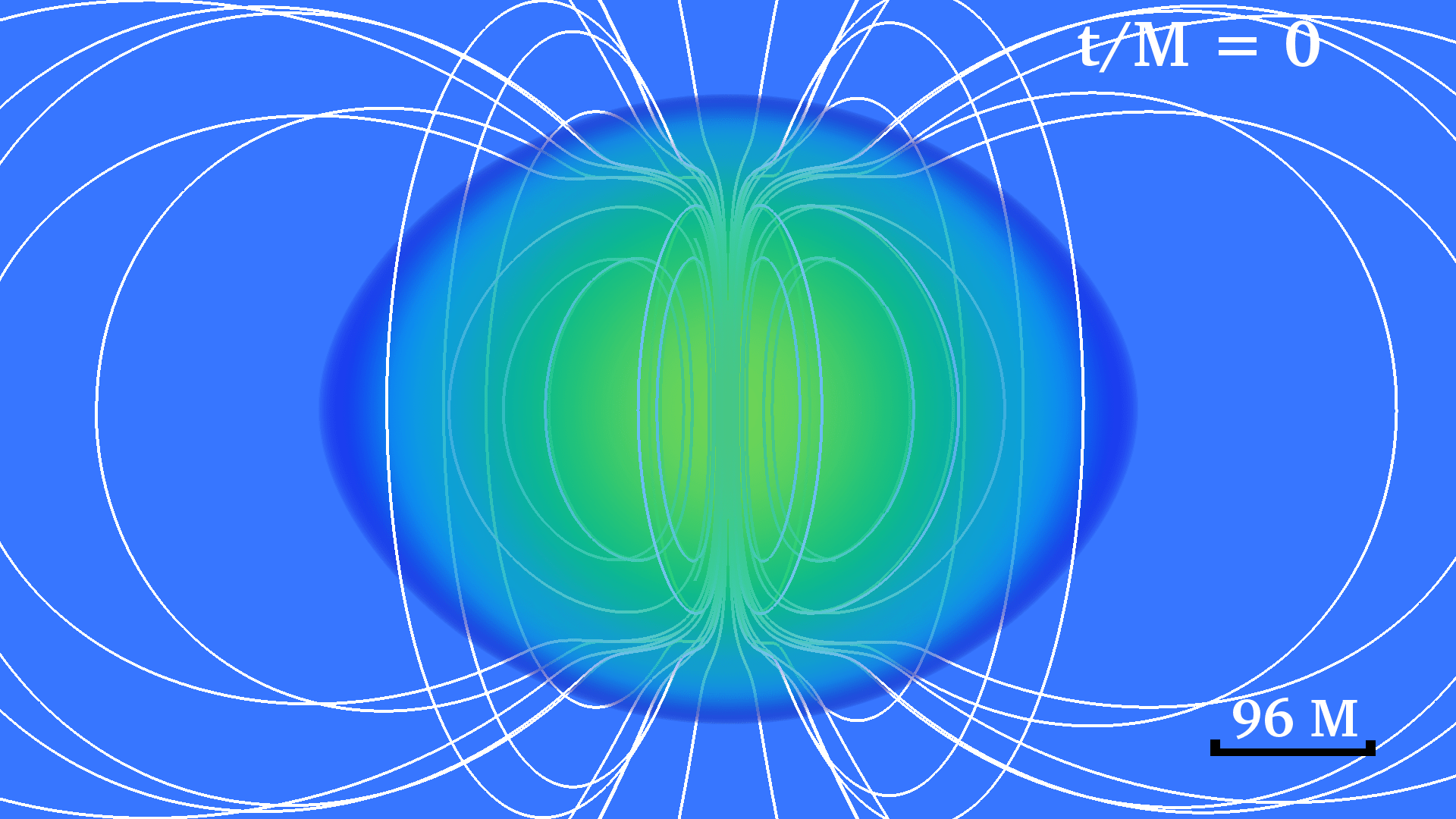}
\includegraphics[height=0.28\textwidth]{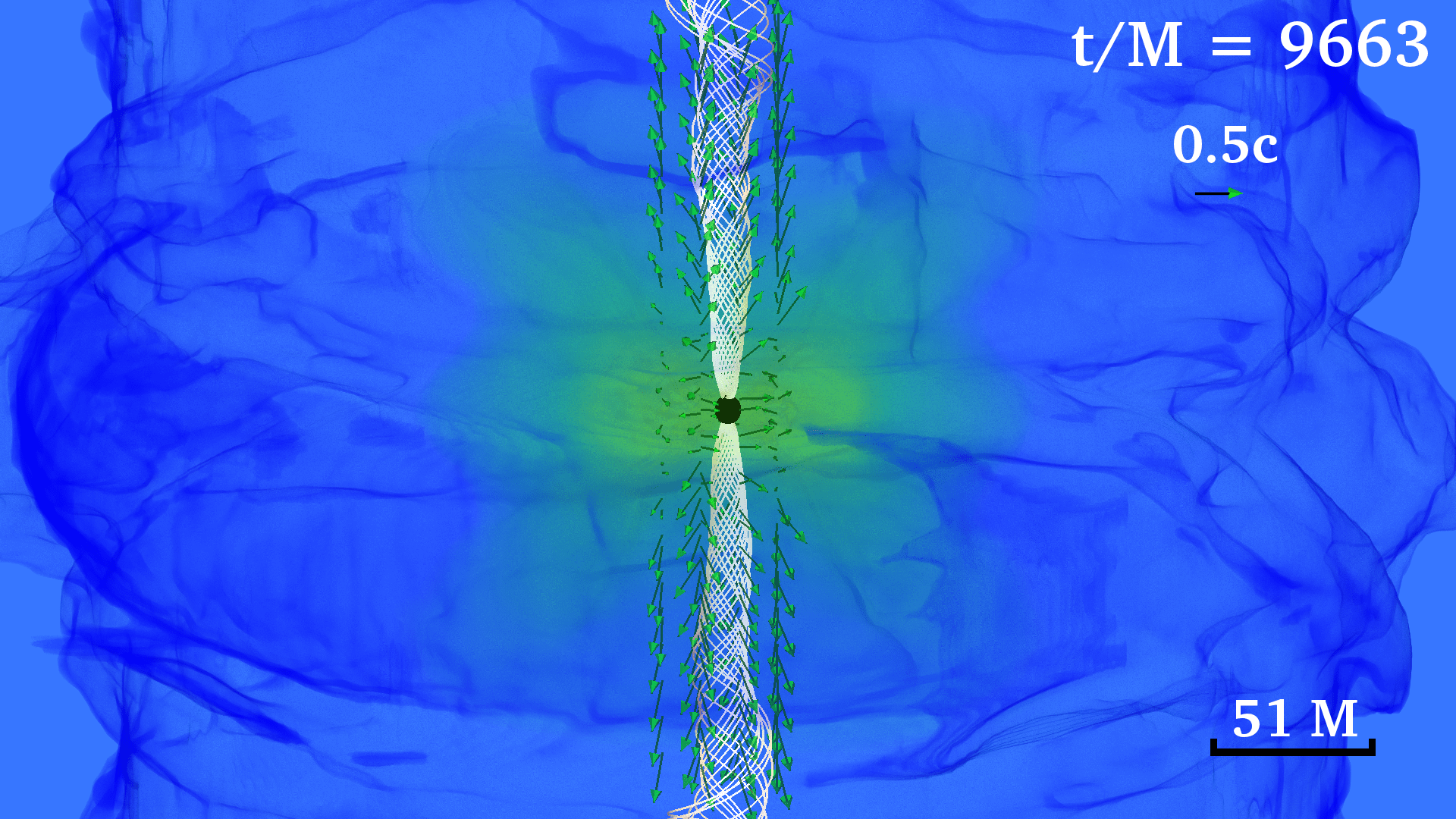}
\includegraphics[height=0.28\textwidth]{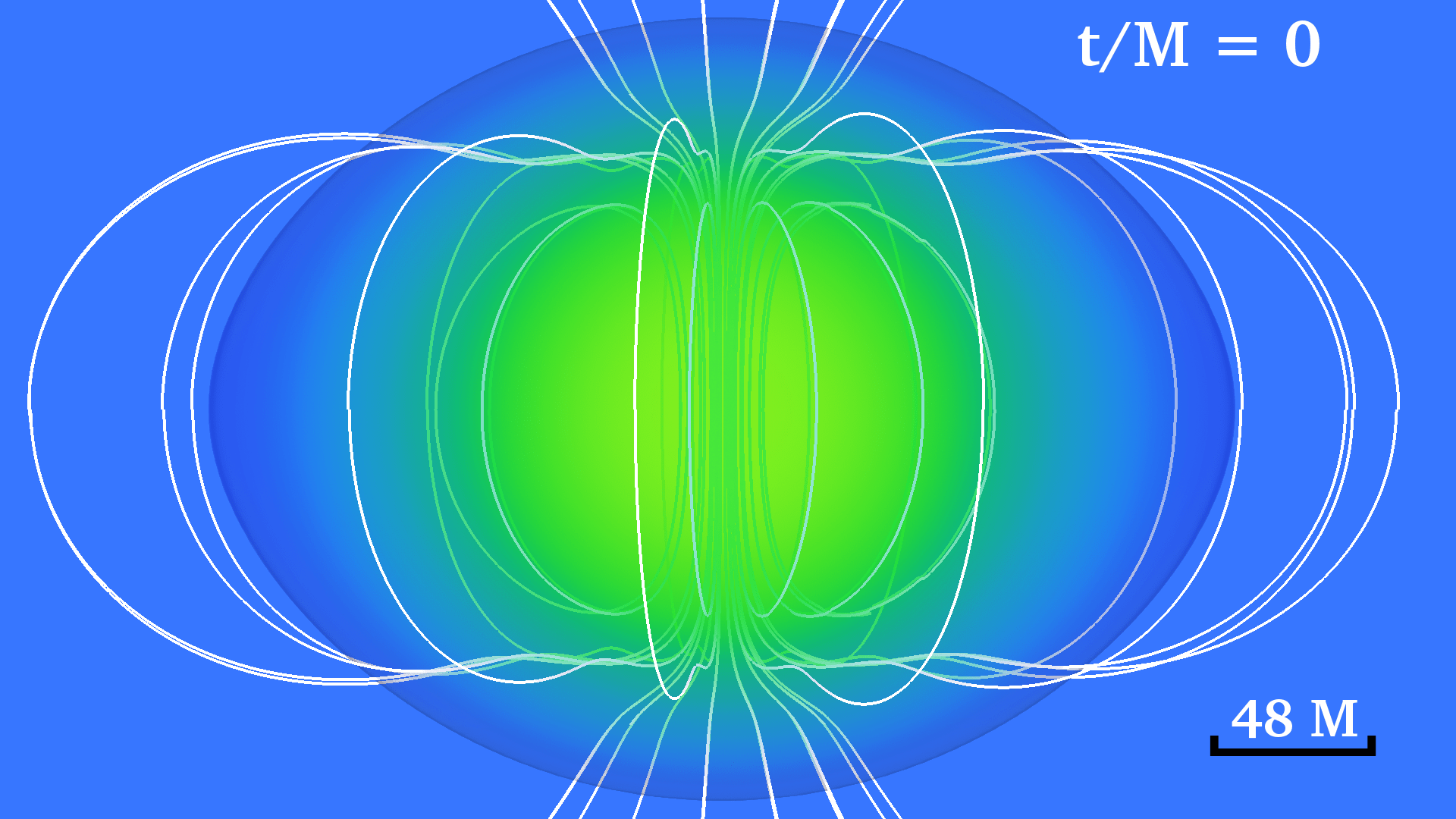}
\includegraphics[height=0.28\textwidth]{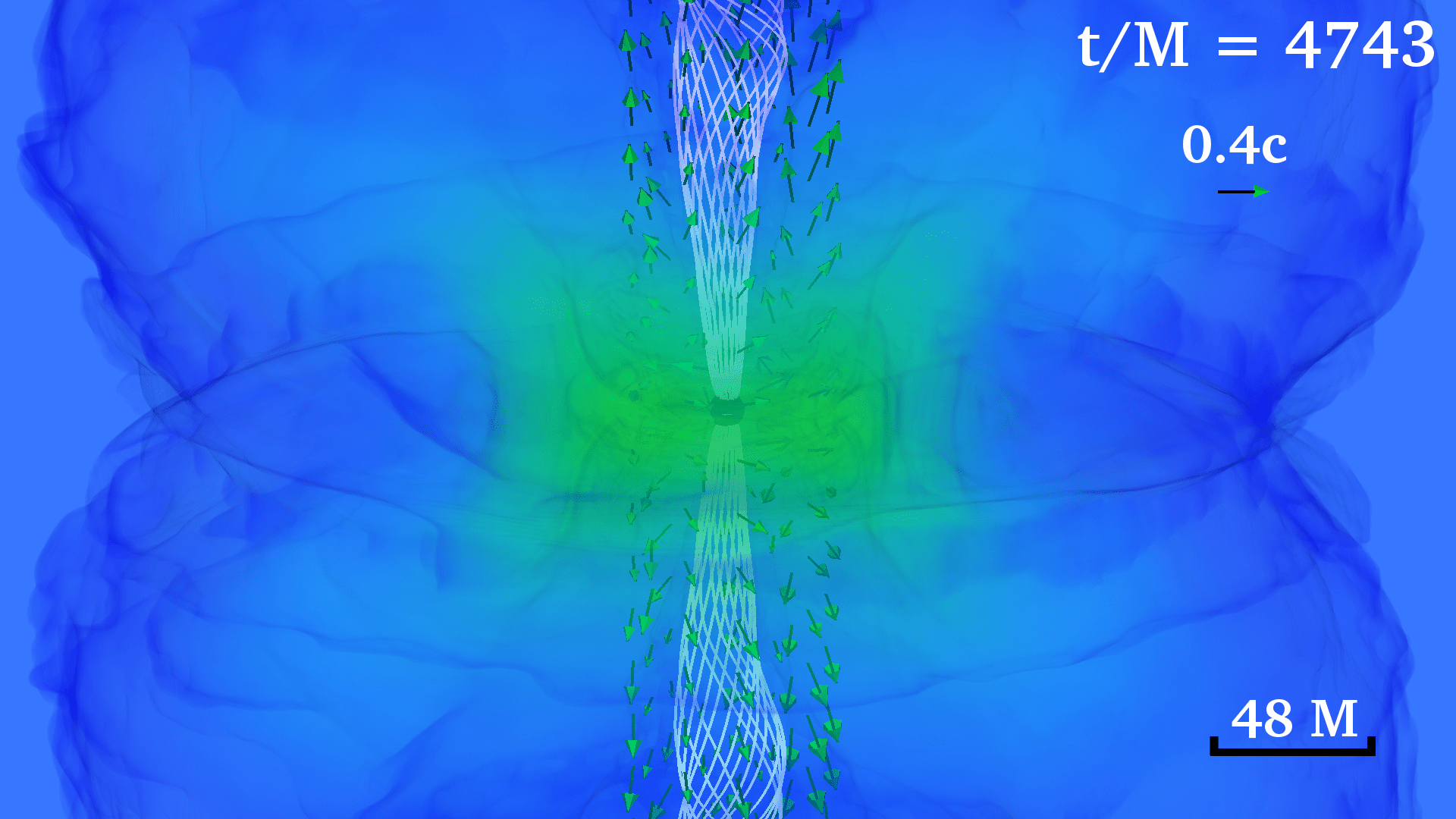}
\includegraphics[height=0.28\textwidth]{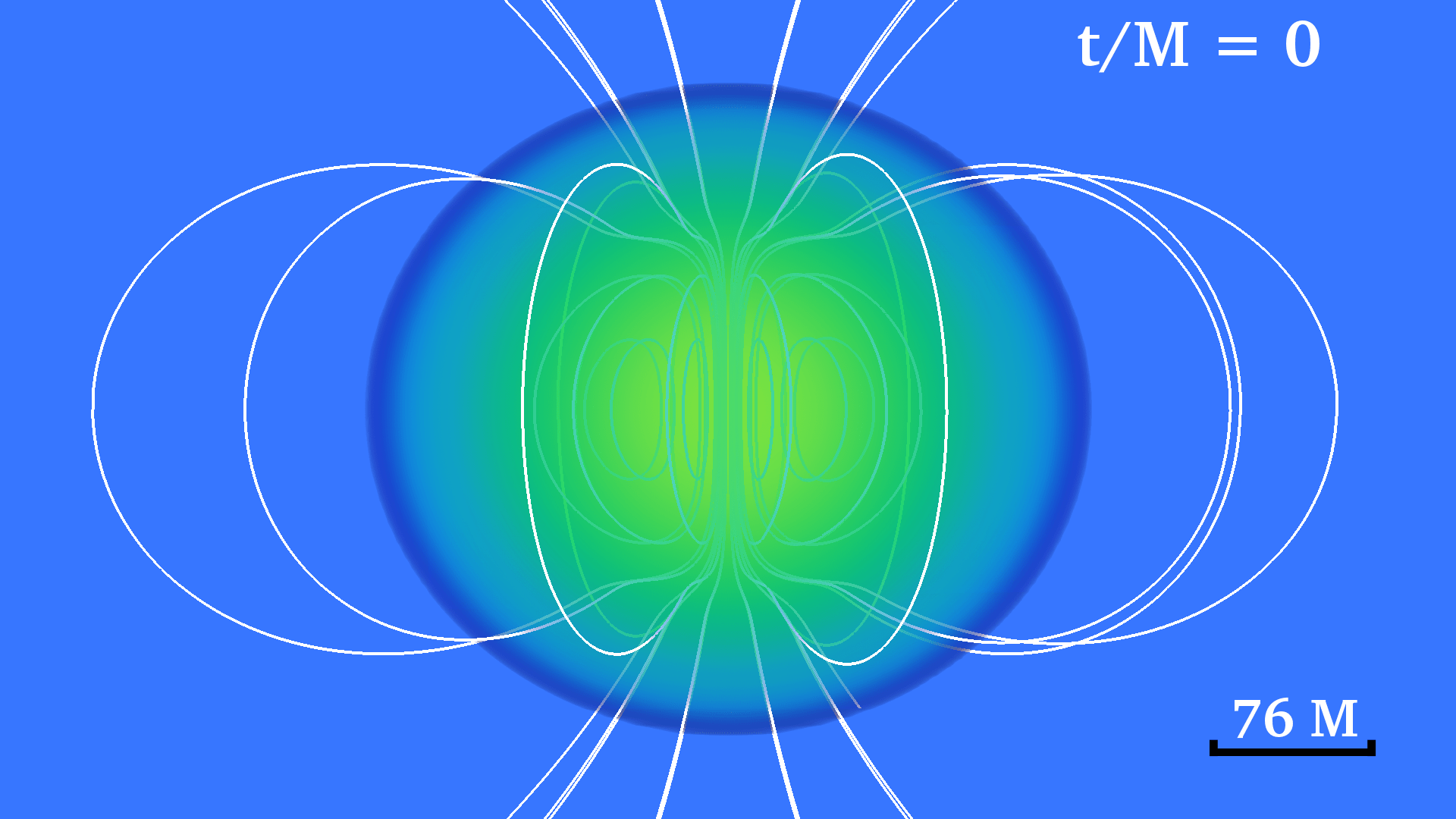}
\includegraphics[height=0.28\textwidth]{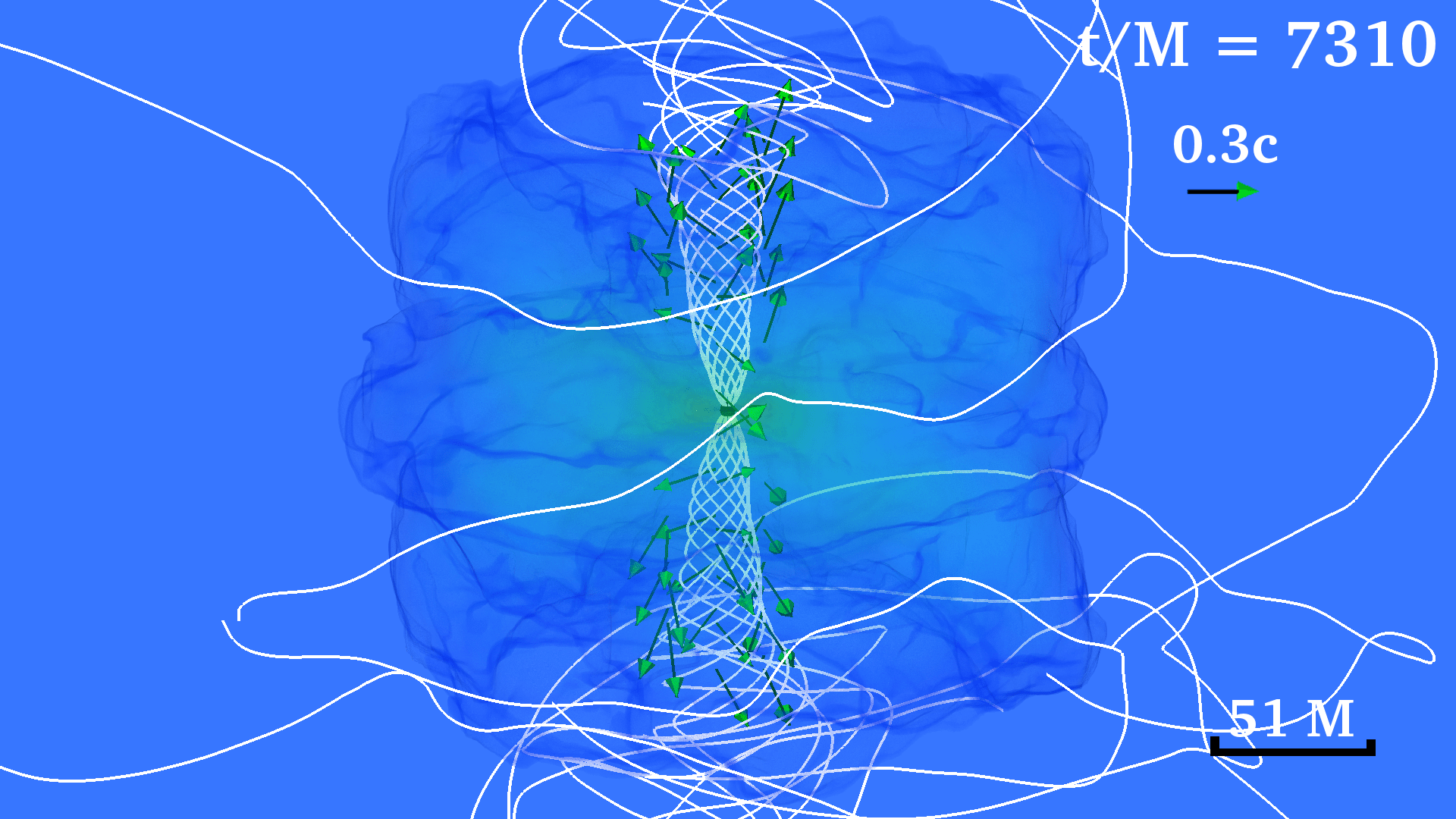}
\caption{\label{fig:initial} 3D volume rendering  of the rest mass density normalized to the corresponding initial
  maximum value  $\rho_{0,max}$ in log scale  (for details see Table~\ref{tab:table1}) for cases n3-EXTINT, n295-EXTINT,
  n29-EXTINT, and n29-EXTINT-0.75SPIN, shown from top to the bottom, respectively. The initial and final configurations for
  these cases are shown in left and right panels, respectively. See Table~\ref{tab:table2} for a summary of global
  parameters describing the final outcome of these cases. Solid lines indicate the magnetic field lines while arrows
  display plasma velocities with length proportional to their magnitude. Here $M=4.9(M/10^6M_\odot){\rm s}=1.47
  \times 10^6(M/10^6M_\odot)$km.}
\end{figure*}

To consider magnetized initial configurations as in~\cite{SunPasRui17}, the stellar models
are endowed with a dynamically unimportant magnetic field as follows:
%
\begin{itemize}
\item {\bf Interior magnetic field case}:
  The star is seeded with a dipole-like magnetic field generated by the vector potential
\cite{EtiLiuPas12}
\begin{align}
  A_\phi^{\rm int} = A_{ b}\,\varpi^2\,\text{max}
  (P-P_{\text{cut}}, 0)^{n_b}\,,
  \label{eq:A_int}
\end{align}
where $A_{ b}$, $P_{\rm cut}$, and
$n_b$ are free parameters that determine  the initial magnetic field strength, its 
confinement and its degree of central condensation. Following~\cite{SunPasRui17}, we set
$P_{\rm cut}=10^{-4}P_{max}(0)$, where $P_{max}(0)$ is the initial maximum value of the pressure,
and $n_b=1/8$. In our models, we choose a value of $A_b$ such as the magnetic-to-rotational-kinetic-energy
ratio $\mathcal{M}/T=0.1$ (see Table~\ref{tab:table1}). As in  standard hydrodynamic
and MHD simulations, we add a tenuous constant--density atmosphere with small rest mass density
$\rho_{0,\,{\rm atm}}=10^{-10}\,\rho_{0,\,\rm max}(0)$, where $\rho_{0,\,\rm max}(0)$  is the
maximum value of the rest mass density of the SMS, to cover the computational grid outside the star.
\item {\bf Interior-Exterior magnetic field case}:  The star is seeded with an interior and exterior
  dipole-like magnetic field generated by the vector potential~\cite{SunPasRui17}
  \begin{equation}
A_\phi =e^{-(r/r_1)^{2p}}A^{\rm int}_\phi+\left(1-e^{-(r/r_1)^{2p}}\right)\,A^{\rm ext}_\phi\,,
  \label{eq:A_int_ext}
  \end{equation}
with
\begin{eqnarray}
  A_\phi^{\rm ext}= \frac{\pi\,\varpi^2\,I_0\,r_0^2}{(r_0^2+r^2)^{3/2}}
  \left[1+\frac{15\,
      r_0^2\,(r_0^2+\varpi^2)}{8\,(r_0^2+r^2)^2}\right]\,,
  \label{eq:A_ext}
\end{eqnarray}
that corresponds to that generated by an interior current loop with radius $r_0$ and
current $I_0$~\cite{Paschalidis:2013jsa,Ruiz:2017inq}. Here, $r^2 = \varpi^2+z^2$ and
the constant $r_0$ is the radius of the current loop that generates the magnetic field
in the stellar exterior. On the other hand, the free constant $r_1$ controls the thickness
of the transition region between the interior and exterior potentials. These parameters,
along with the current loop $I_0$ and the free parameter $p$, determine the strength of
the magnetic field. Following~\cite{SunPasRui17}, in all models
listed in Table~\ref{tab:table1} we choose $P_{\rm cut}=10^{-4} P_{\rm max}$ and
$I_0=7.35\times 10^{-3}$. In the $n=3.0$ SMS model, we set $r_0 \approx 2.2 M$, and
$r_1\approx 240M$. In the $n=2.95$ model,
we set $r_0 \approx 0.6 M$, and  $r_1\approx 120M$, and, finally, in the $n=2.9$ model,
we set $r_0 \approx 0.6 M$, and $r_1\approx 120M$. In all cases we set $p = 2$.  The above
choices yield a magnetic field in the bulk of the star similar to that in the interior
case\cite{SunPasRui17}. Finally, we set an initial low and variable density atmosphere in
the stellar exterior such that the gas-to-magnetic-pressure
ratio is $0.01$ which allows us to evolve reliably the magnetic field outside the star and
mimic a force-free external environment~\cite{PasRuiSha15,RuiLanPas16}. The left top panel
in Fig.~\ref{fig:Evolution_diff} and the left column in Fig.~\ref{fig:initial} display
the initial magnetic field configurations of the models listed (see~Table~\ref{tab:table1}).
\end{itemize}
Since we are interested in the stellar collapse epoch and the subsequent evolution, we initially deplete
the pressure by $1\%$ as in~\cite{LiuShaSte07, SunPasRui17} to
trigger stellar collapse. Table~\ref{tab:table1} summarizes the key initial
parameters of these models.  Unless otherwise noted, the initial configuration corresponds to a uniformly
rotating SMS spinning at the mass-shedding limit, close to the onset of general relativistic radial instability.
So, for example, the model denoted as n29-EXTINT-DIFF corresponds to an $n=2.9$ differentially rotating star
endowed with a magnetic field that extends from the stellar interior to the exterior, while the model denoted
as n29-HYD corresponds to the $n=2.9$ uniform rotating star spinning at the mass-shedding limit without any
magnetic field.
%
%
\subsection{Grid Structure}
\label{subsec:grid}
During the collapse, the size of the star changes in many orders of magnitude from some
hundreds of $M$ to a few $M$ (see Table~\ref{tab:table1}). Hence, to reliably evolve the SMS,
high-resolution refinement levels need to be added on the base levels as the star size shrinks.
Following~\cite{ShiSha02,LiuShaSte07,SunPasRui17}, we begin
the numerical evolution of the models listed in Table~\ref{tab:table1} with one set of five nested
refinement levels centered at the star and differing in size and resolution by factors of two. Reflection
symmetry across the equatorial plane is imposed to save computational resources.  The resulting number
of grid points per level is  $N=N_x\times N_y\times N_z\geq 120^2\times 60$, where $N_i$ is the number
of grids points along the $i$--direction. During the evolution, a new refinement
level is added each time  the central density increases by roughly a factor of three. The new level
has half the grid spacing of the previous innermost level with same number of grid points. Such
a procedure is repeated five and six times for the $n=3$ purely hydrodynamic and GRMHD evolutions,
respectively, and  four times for the other cases (see Table~\ref{tab:resolution}). The highest
resolution on our grids is similar to that used in~\cite{LiuShaSte07,SunPasRui17}. Note that
the main purpose of applying higher resolution is to  accurately evolve  the low-density, force-free
environments that emerge above the black hole poles.
%
%
\begin{table}[t]
  \caption{\label{tab:resolution} Grid structure for all cases listed in Table~\ref{tab:table1}.
    The computational mesh consists of one set of $j$-nested AMR  grids centered at the start, in
    which equatorial symmetry is imposed. Here $j=5,\cdots,\,\rm{level}_{\rm max}$ denotes the
    number of AMR grids during a given evolution epoch, and $\rm{level}_{\rm max}$ is the maximum
    number of AMR grids at the end of the simulations. Each case begins with a set of $j=5$-AMR grids,
    and we add a new refinement level every time the maximum value of the rest-mass density increases
    by a factor of three. The finest level for a given set of $j$-nested grids is denoted by
    $\Delta x_{\rm min}$.  The grid spacing of all other levels is $2^{l-1}\,\Delta x_{\rm min}$,
    where $l=1,\cdots,\,j$, is the level number  such that $l=1$ corresponds to the coarsest
    level. The half-side length of the outermost AMR boundary  is given by the first number in
    the grid hierarchy.}
\begin{ruledtabular}
\begin{tabular}{ccccc}
Case& $\Delta x_{\rm min}$  & $\rm level_{\rm max}$ & Grid hierarchy& \\
\hline
n3-HYD            &  $1.36M/2^{j-5}$ &10 & $1312M/2^{l-1}$ \\
n3-INT            &  $1.36M/2^{j-5}$ &11 & $1312M/2^{l-1}$ \\ 
n3-EXTINT         &  $1.36M/2^{j-5}$ &11 & $1312M/2^{l-1}$ \\ 
n295-EXTINT       &  $0.4M/2^{j-5}$  & 9 & $728M/2^{l-1}$   \\
n29-HYD           &  $0.48M/2^{j-5}$ & 9 & $454M/2^{l-1}$ \\
n29-INT           &  $0.48M/2^{j-5}$ & 9 & $454M/2^{l-1}$ \\
n29-EXTINT        &  $0.48M/2^{j-5}$ & 9 & $454M/2^{l-1}$ \\
n29-EXTINT-0.75SPIN &  $0.48M/2^{j-5}$ & 9 & $458M/2^{l-1}$ \\
n29-EXTINT-DIFF   &  $0.40M /2^{j-5}$ & 9 & $515M/2^{l-1}$   \\
\end{tabular}
\end{ruledtabular}
\end{table}
%
\begin{figure}[h]
  \includegraphics[width=0.41\textwidth]{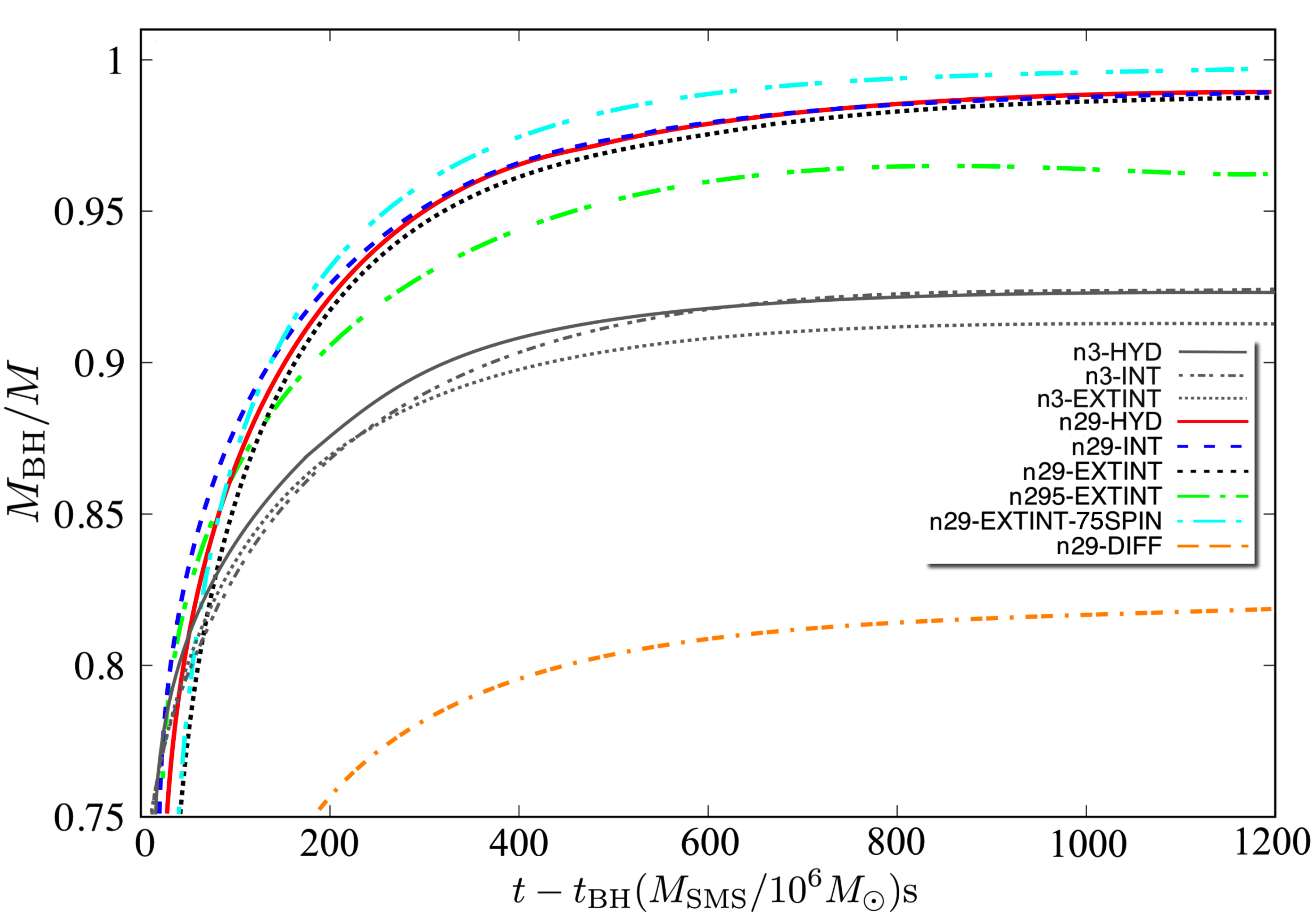}
  \includegraphics[width=0.41\textwidth]{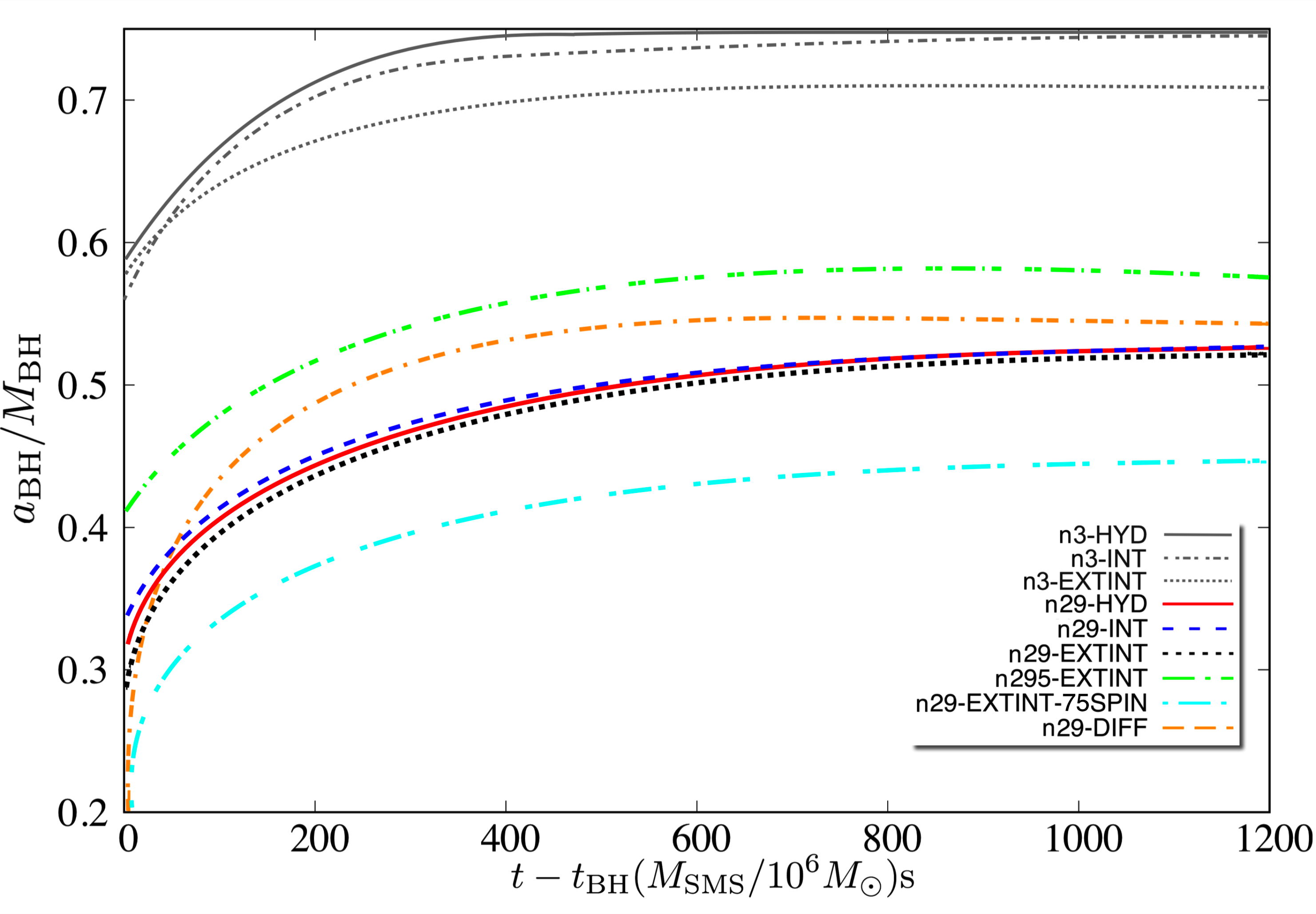}
  \includegraphics[width=0.415\textwidth]{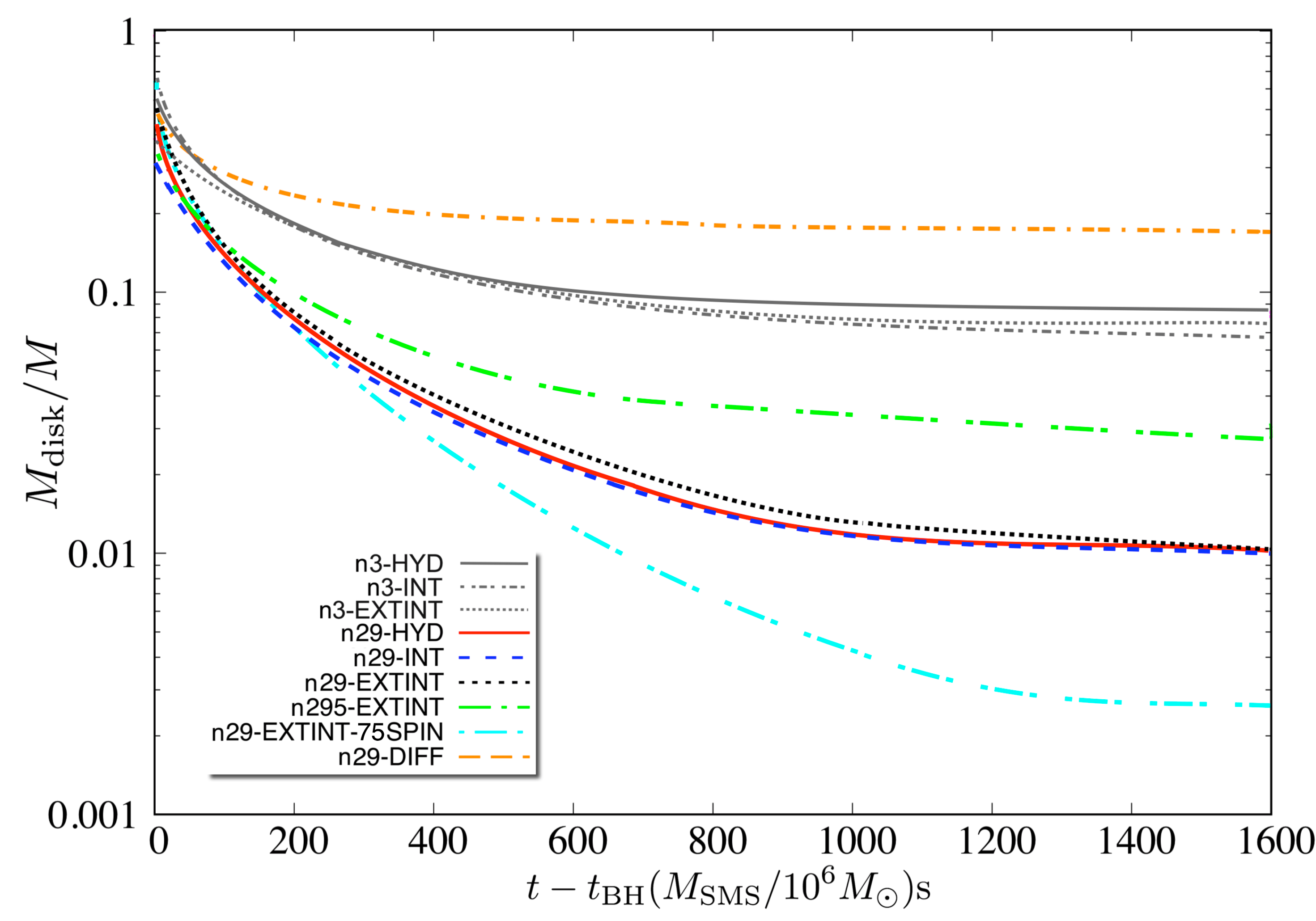}
   \caption{\label{fig:BH_m_a}
     Dependence of the black hole mass (top panel), black hole dimensionless spin parameter (middle panel), and
     the accretion mass disk (bottom panel) on different EOSs, magnetic field configurations, and rotation profiles
     for models in Table~\ref{tab:table1}. The  mass of the black hole remnant, and hence the mass of the disk, is
     sharply sensitive to changes in the EOS as well as the initial rotation profile of the SMS.}
\end{figure}
%
\begin{figure}[h]
  \hspace*{-0.35cm}\includegraphics[scale=0.09]{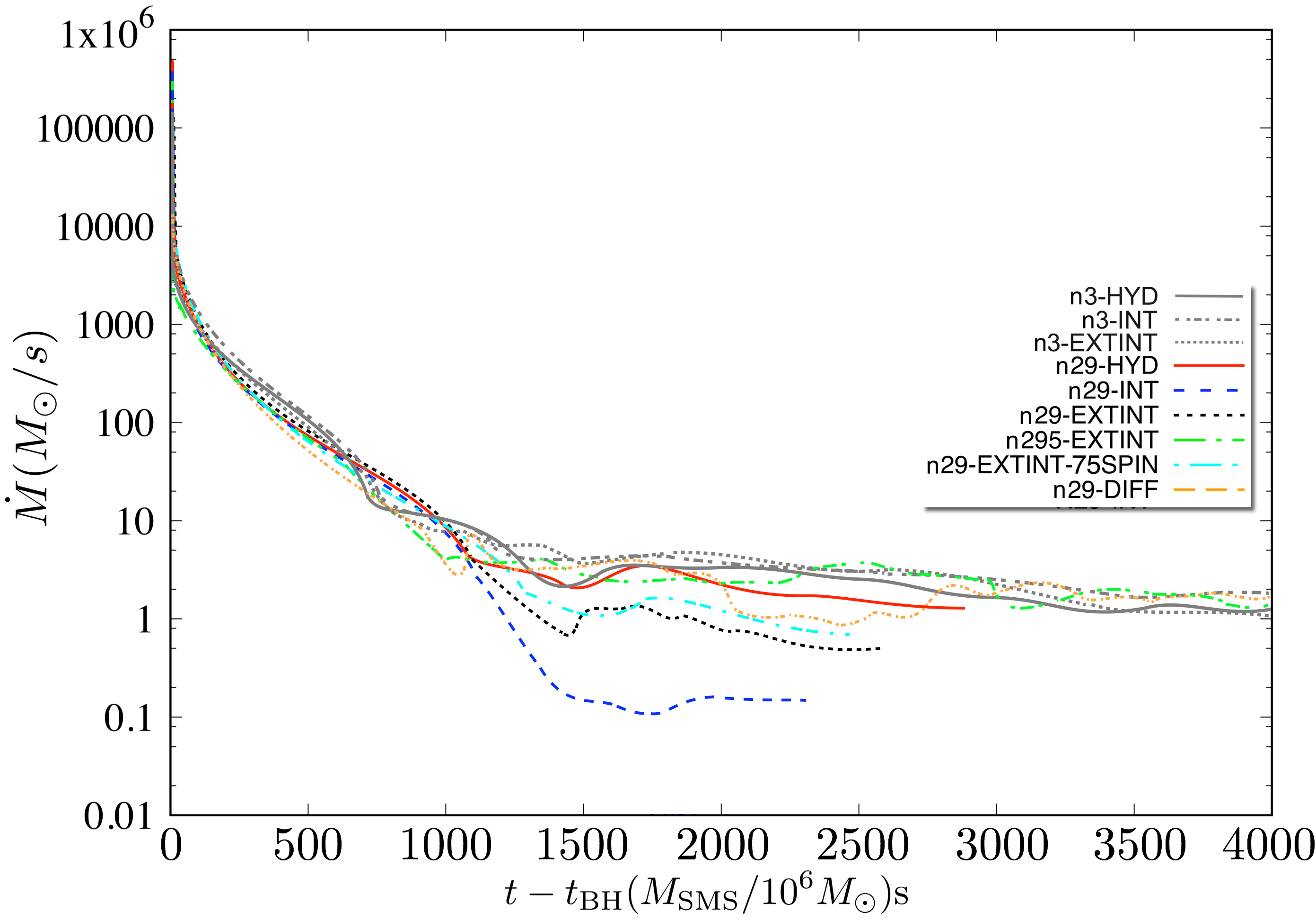}
  \caption{\label{fig:Mdot} 
    Rest mass accretion rate $\dot{M}$ vs. time for cases listed in Table ~\ref{tab:table1}.
    Notice that the time has shifted to the black hole formation time and it is normalized
    to $(M/10^{6} M_{\odot})$.}
\end{figure}
%
%
\subsection{Diagnostics}
\label{subsec:diag}
During the evolution, we monitor the normalized Hamiltonian  and momentum constraints calculated
by Eqs.(40)-(43) in~\cite{Etienne:2007jg}. In all cases displayed in table~\ref{tab:table1},
the constraint violations remain below $\sim 0.01$ throughout the whole evolution.  We use a modified
version of the {\tt Psikadelia} thorn to extract GWs using the Weyl scalar $\Psi_4$ and computed
the total energy radiated by gravitational waves; this routine uses a  $s=-2$ spin-weighted spherical
harmonics decomposition~(for details see~\cite{RuiAlcNun07}). To further validate our numerical results,
we verify the conservation of the total mass $M_{\rm int}$ and the total angular momentum
$J_{\rm int}$ computed through Eqs.(9)-(10) in~\cite{EtiLiuSha08},
which coincides with the ADM mass only at spatial infinity. In all cases we find that 
both the interior mass and the interior angular momentum calculate at large but finite radius
deviate from their initial values by $\lesssim 1\%$, which is manly due to numerical dissipation.
Notice that in the above calculation we  take into account the energy and angular momentum carried
away by gravitational radiation, which is $\lesssim 10^{-4}\%$, the mass and angular momentum loss
through EM radiation, computed via Eq.(7) in~\cite{SunPasRui17}, as well as the escaping matter,
which computed as $M_{\rm esc} =\int_{-u_0>1} \rho_*\,d^3x$ with $\rho_*=-n_\mu\,\rho_0\,u^\mu$,
where $\rho_0$, $u^\mu$ and $n_\mu$ are the rest-mass density, 4-velocity, and the future-directed
unit normal to the time slice. respectively.

Finally, we use the {\tt AHFinderDirect} thorn~\cite{Thornburg:2003sf} to locate the apparent horizon,
as well as the isolated horizon formalism to estimate the spin and mass of the
black hole via Eqs.(25) and (27) in~\cite{DreKriSho02}. 
%
%
\section{Results}
\label{sec:results}
%
%
\subsection{Overview}
Following the initial pressure depletion, the bulk of our SMS models begin to undergo nearly homologous
collapse. Regardless of the different characteristic masses (or polytropic index $n$), the magnetic field
configuration, or the stellar rotation law, the gas falls inward, forming a dense core that  eventually
collapses to a black hole. Following the catastrophic collapse, the black hole captures in all the low-angular
momentum gas from the inner layers of the SMS. The high-angular momentum gas in the outer layers spirals around
the black hole as it falls inward and is ultimately held back by a centrifugal barrier. Eventually, a reverse
shock is formed which induces an outflow~(see e.g.~\cite{LiuShaSte07,SunPasRui17}). During this epoch, the
frozen-in magnetic field winds up (see right top and middle panels of Fig.~\ref{fig:Evolution_diff}),
and the magnetic pressure grows. The magnetorotational-instability (MRI) develops in the disk. We resolve the
MRI according to $\lambda / \Delta \approx 10 -20 $~\cite{Gold:2013zma}. Here $\lambda$ is the wavelength of
the fastest-growing MRI mode and $\Delta$ is the grid spacing.  Once the magnetic pressure above the black hole
poles is sufficiently large (i.e. $B^2/8\pi\rho_0\gtrsim 1$), a collimated outflow is driven along the polar axis of
black hole, and an incipient jet is launched~\cite{SunPasRui17} (see bottom panels of Fig.~\ref{fig:Evolution_diff}
and right column of Fig.~\ref{fig:initial}).
%
\begin{table*}[!ht]
  \caption{\label{tab:table2} Summary of key results. Here $t_{\rm BH}$ denotes the black hole formation time, 
    $M_{\text{BH}}$  and $a_{\text{BH}}/M_{\text{BH}}$ denote the mass of the black hole and the dimensionless spin
    once the system has settled down, respectively, $R_{disk}$ and $M_{disk}$ are the outer edge and the
    mass of the accretion disk, $\dot{M}$ is the accretion rate roughly after  $t-t_{BH}\sim 1.8\times 10^3M
     \sim 8.8\times 10^3(M/10^6M_\odot)$s, $t_{jet}$ is the launching jet time after black hole formation,
    $\tau_{disk}\equiv M_{disk}/\dot{M}$ is disk lifetime, $(b^2/2\rho_0)_{| \rm pole}$ and
     $L_{\rm EM}$ are the force-free parameter above the black hole pole and the Poynting electromagnetic luminosity
     driven by the incipient jet, respectively,
     which are time-averaged over  $t\approx 200M\sim 10^3 (M/10^6M_{\odot})$s after jet launching.
    The quantities $t_{BH}$, $t_{jet}$, and $\tau_{disk}$ are normalized by $(M/10^6M_\odot)$.
  }
  \begin{ruledtabular}
\begin{tabular}{cccccccccccccc}
Case&$t_{\text{BH}}$(s) & $M_{\text{BH}}/M$ & $a_{\text{BH}}/M_{\text{BH}}$ & $R_{disk}/M$ &$M_{disk}/M$ &$\dot{M}(M_{\odot}/s)$ & $t_{jet}$(s) & $\tau_{disk}$(s) & $(b^2/2\rho_0)_{| \rm pole}$ &$L_{\rm EM}(\rm erg/s)$\\
\hline
n3-HYD            &  $1.40\times 10^5$   & 0.91  & 0.75  & $  95$ & $9.0\%$   &$  1.0$     & -                 & $9.0\times 10^4$&-            & -           \\
n3-INT            &  $1.48\times 10^5$   & 0.92  & 0.74  & $  90$ & $6.0\%$   &$  1.2$     & $2.7\times 10^3$  & $5.0\times 10^4$&$  25$       &$10^{50.6}$  \\
n3-EXTINT         &  $1.53\times 10^5$   & 0.92  & 0.68  & $  95$ & $7.0\%$   &$  1.1$     & $2.2\times 10^3$  & $7.2\times 10^4$&$  300$      & $10^{52.5}$ \\ 
n295-EXTINT       &  $4.48\times 10^4$   & 0.96  & 0.58  & $  75$ & $3.0\%$   &$  1.2$     & $1.5\times 10^3$  & $2.4\times 10^4$&$  100$      &$10^{52.3}$  \\
n29-HYD           &  $2.06\times 10^4$   & 0.99  & 0.53  & $  60$ & $1.1\%$   &$  1.0$     & -                 & $1.0\times 10^4$&-            &  -          \\
n29-INT           &  $2.13\times 10^4$   & 0.99  & 0.53  & $  55$ & $1.1\%$   &$  0.4$     & -                 & $1.0\times 10^4$&$< 10^{-4}$  &  -          \\
n29-EXTINT        &  $2.12\times 10^4$   & 0.99  & 0.52  & $  55$ & $1.5\%$   &$  0.8$     & $1.2\times 10^3$  & $1.8\times 10^4$&$  100$      & $10^{52.1}$ \\
n29-EXTINT-0.75SPIN &  $3.25\times 10^4$   & 0.99  & 0.45  & $  55$ & $0.3\%$   &$  1.5$     & $1.7\times 10^3$  & $1.0\times 10^4$&$  60$       & $10^{51.5}$ \\
n29-EXTINT-DIFF   &  $1.26\times 10^5$   & 0.82  & 0.54  & $  60$ & $18.0\%$  &$  2.0$     & $1.4\times 10^3$  & $9.0\times 10^4$&$  300$      & $10^{53.5}$ \\
\end{tabular}
\end{ruledtabular}
\end{table*}
%
\begin{table*}[t]
  \caption{\label{tab:comparison} Comparison of black hole and disk parameters from cases in Table \ref{tab:table2}
    (bold) with the semi-analytic and numeric results in previous studies for critical collapse at mass-shedding
    different EOS (characteristic mass) and magnetic fields. Here ``H", ``I", and ``E+I" represent no magnetic field,
    interior magnetic field, and exterior plus interior magnetic field, respectively.}
\begin{ruledtabular}
\begin{tabular}{cccccccccccc} 
  \multirow{2}{*}{$n$ }  &
  \multicolumn{3}{c}{\underline{\hspace{0.95cm} $M_{BH}/M$ \hspace{0.85cm}}} &
  \multicolumn{3}{c}{\underline{\hspace{0.85cm} $a_{BH}/M_{BH}$ \hspace{0.75cm}} }& 
  \multicolumn{3}{c}{\underline{\hspace{0.85cm} $M_{disk}/M$\hspace{0.85cm}}}\\
  $(M_{\star})$& H & I & E+I \hspace{0.35cm}& H & I & E+I\hspace{0.05cm} & H & I & E+I\\
  \hline\hline
  3.00   &  0.89\footnote{Table 2 in~\cite{Sha04}, fully analytic}  & 0.95 \footnote{GRMHD simulation by
    \cite{LiuShaSte07} (model S1)}  &\textbf{0.94} \hspace{0.35cm}& 0.60  & 0.70& \textbf{0.68} & $11.0\%$   &  $6.0\%$
  &  $\mathbf{7.0\%}$                                                              \\
  ($\gtrsim 10^6M_{\odot}$)&0.87\footnote{Table 2 in~\cite{Sha04}, analytic, using critical configuration in~\cite{BauSha99}}
  & 0.95\footnote{GRMHD simulation by ~\cite{LiuShaSte07} (model S2)} &\hspace{0.35cm}&0.71 & 0.68&  & $13.0\%$   &    $6.0\%$\\ 
  &0.90\footnote{GR hydrodynamic simulation by ~\cite{ShiSha02}}   & \textbf{0.92}
  &\hspace{0.35cm}&0.75 & \textbf{0.64}& & $10.0\%$   &   $\mathbf{6.0\%}$ &          \\ 
  
  &0.90\footnote{GR hydrodynamic simulation by ~\cite{LiuShaSte07} (model S0)}  & &&  0.70 & & & $7.0\%$ & &
  \\
  &\textbf{0.91} & & &\textbf{0.75} && &$\mathbf{9.0\%}$ &&\\
  \hline
   2.95  &0.97$^{\rm a}$   & & \textbf{0.96}  \hspace{0.25cm}&0.52 & & \textbf{0.58} & $2.9\%$ & &   $\mathbf{3.0\%}$  	      \\
    $(\sim 10^5M_{\odot})$  &  &  & &     	      \\
  \hline
  2.90 & 0.99$^{\rm a}$  & \textbf{0.99}  & \textbf{0.99}  \hspace{0.25cm} &0.45 & \textbf{0.53} & \textbf{0.52}& $1.1\%$ &    $\mathbf{1.1\%}$    &    $\mathbf{1.5\%}$     \\
  
  $(\sim 10^4M_{\odot})$ &0.99 \footnote{Table 2 in ~\cite{CooShaTeu94} by setting $R_p/R_e \approx 2/3$}   &   & & 0.53 & &	& $1.4\%$   &    & \\
  &\textbf{0.99}   & &  &   \textbf{0.53}   & & &   $\mathbf{1.4\%}$
  & 	  &                                                              \\
\end{tabular}
\end{ruledtabular}
\end{table*}
%
\subsection{Effects of different mass scale}
\label{sec:effects_EOS}
Semianalytic calculations of marginally unstable, uniformly rotating and axisymmetric
SMS spinning at the mass-shedding limit in~\cite{Sha04}, and numerical calculations
(see e.g.~\cite{BauSha99,Shibata:2016vzw,LiuShaSte07,SunPasRui17}) of SMSs supported by
thermal radiation pressure with $\Gamma \approx 4/3$ suggested that the final parameters
that characterize the BH-accretion disk remnant depend strongly on $\Gamma -4/3 \ll 1$, or $n-3 \ll 1$.
We consider  the evolution of marginally unstable SMSs spinning at the mass-shedding limit
described by a polytropic  EOS with $n\in \{3.0,\,2.95,\,2.90\}$ (see Table~\ref{tab:table1}),
which characterize masses of $M_{\star} \in \{\gtrsim 10^6,\,10^5,\,10^4\}M_{\odot}$ respectively,
supported by radiation plus gas pressure. Although the different $n$ characterize different mass
scales, we nevertheless scale our numerical results in units of $10^6 M_{\odot}$ for convenient
comparisons. 

The stiffer the EOS (the smaller the characteristic mass $M_{\star}$), the more compact the critical
configuration and, hence, the shorter the black formation time.
We observe that in the most massive SMS models with $n=3$, an apparent horizon (AH) forms by $t\approx
3.0\times 10^{4}M\sim 1.5\times 10^{5}(M/10^6M_\odot)$s~\cite{SunPasRui17}. For the
$n=2.95$ SMS, the AH forms by $t\approx {9.08}\times 10^{3}M\sim {4.48}\times 10^{5}
(M/10^6M_\odot)$s, while for the models with $n=2.9$, the horizon appears about
$t\approx 4.2\times 10^{3}M\sim 2.1\times 10^{4}(M/10^6M_\odot)$s.
Regardless of EOS, in all cases listed in table~\ref{tab:table1},
we observe that following the high accretion episode, both the mass and the spin of the black hole
rapidly grow for about $t-t_{BH}\approx 400M\sim 1.8\times 10^3(M/10^6M_\odot)$s until reaching
quasistationary state values. Fig.~\ref{fig:BH_m_a} shows the dependence of these quantities on the
polytropic index $n$ (see Table~\ref{tab:table2} for details).  For models with the smallest masses
(stiffer EOS, $n \rightarrow 2.9$),
essentially all the mass and the angular momentum of the progenitor are swallowed by the black
hole during the high accretion episode, leaving  only a tenuous cloud of gas to form the accretion
disk. We find that only $\sim 1\%$ of the SMS rest mass ends up in the disk, and the final spin of the
black hole remnant is $a/M_{BH}\approx 0.53$ which is approximately the initial angular momentum
of the SMS. On the other hand, as the characteristic mass becomes greater (softer EOS, $n \rightarrow 3$),
the initial SMS configuration becomes less
compact (see table~\ref{tab:table1}), allowing more gas to be sufficiently far from the 
final BH innermost stable circular orbit (ISCO) allowing for a higher mass of the disk. For $n=2.95$,
we find that around $\sim 3\%$ of the SMS exists in the disk, but
for $n=3.0$ it can be as much as $\sim9\%$ of the SMS rest mass~\cite{SunPasRui17}. The spin of the
black hole for these cases is $a/M_{BH}\approx 0.58$, and $a/M_{BH}\approx 0.7$, respectively. Although
the softer EOS produces larger $M_{\rm BH}$ and $a_{\rm BH}/M_{\rm BH}$, only the mass of the black hole
seems to be sharply dependent on the polytropic index $n$. Note that the above results are consistent
with the previous simulations of the collapse of $n \approx 2.98$ SMS models reported in
\cite{Uchida:2017qwn} that account for nuclear burning, for which the mass of the disk is $\lesssim 5\%$,
a value that lies between our $n=3$ and $n=2.95$ SMS models, as expected.

Finally,  we compare our numerical results with the semianalytic predictions for the collapse of critical
configurations uniformly rotating at mass-shedding in ~\cite{CooShaTeu94, BauSha99,Sha04} and previous GR
hydrodynamic simulations in ~\cite{ShiSha02} and  GRMHD simulations in~\cite{LiuShaSte07}.
As it can be seen in Table~\ref{tab:comparison}, the previous theoretical predictions and numerical
calculations are consistent with the results of our simulations for the mass of BH, the dimensionless
spin of BH, and the disk mass for all three polytropic indices (characteristic mass scales). 
%
\subsection{Effects of different rotation law}
\label{sec:effects_Rotation}
If the turbulent
viscosity is low, uniform rotation may not be enforced during stellar evolution,
and hence the star may be differentially rotating when it collapses to a black hole (see e.g.
\cite{NewSha00a}). Since  the angular momentum of the outer layers of the collapsing star will be
conserved, the fate of the remnant black hole-disk will depend on the initial rotation law profile
of the SMS~\cite{BodOst73,Tas78,NewSha00a,Shapiro:2003xe}.  Fig.~\ref{fig:BH_m_a} displays the evolution
of the mass and the spin  of the black hole remnant, as well as the rest mass fraction $M_0$ outside the
AH computed as $M_0=\int \rho_*\,d^3x$ for models listed in table~\ref{tab:table1}. Here we focus on
the  three different $n=2.9$ SMS models that:  $a)$ uniformly rotate at the  mass-shedding limit,
$b)$ uniformly rotate at 75\% of the mass-shedding limit, and $c)$ differentially rotate with an
initial rotation profile given by Eq.~(\ref{eq:rotation_pro}).

Following the initial pressure depletion, the SMS models contract and form a central dense core
that undergoes collapse. Unlike the uniform  rotation  models, in which an AH forms approximately
at $t\lesssim 6600M\sim 3.2\times 10^4(M/10^6M_\odot)$s, the differential rotation profile provides
centrifugal support against collapse. However, over a secular time $t\lesssim 2 \times 10^4 M\sim
10^5(M/10^6M_\odot)$s, turbulent viscosity (in our case magnetic viscosity) transports the angular
momentum outward pushing out the external layers of the star and driving the inner core toward uniform
rotation. As the total rest mass of the core exceeds the maximum
value allowed by uniform rotation, it eventually collapses. The black hole horizon appears by around $t\approx
2.6\times 10^3M\sim 1.3\times 10^5(M/10^6M_\odot)$s, which is similar to that in the case of a less
centrally condensed $n=3$ SMS (see table~\ref{tab:table2}). The bottom panel of Fig.~\ref{fig:BH_m_a}
shows the fraction of the rest mass that  wraps around the black hole to form the accretion disk.
Notice that in the differentially rotating case about $\sim 18\%$ of the initial rest mass
of the star forms the disk, while only $\sim 1\%$ and  $\sim 0.3\%$ of the rest mass contributes to the
disk in the uniformly rotating mass-shedding case, and in the $\alpha =0.75$--uniformly
rotating case, respectively.
%
\subsection{Effects of the magnetic field: Jets}
\label{sec:effectsjet}
\begin{figure}[h]
  \hspace{-0.35cm}\includegraphics[scale=0.09]{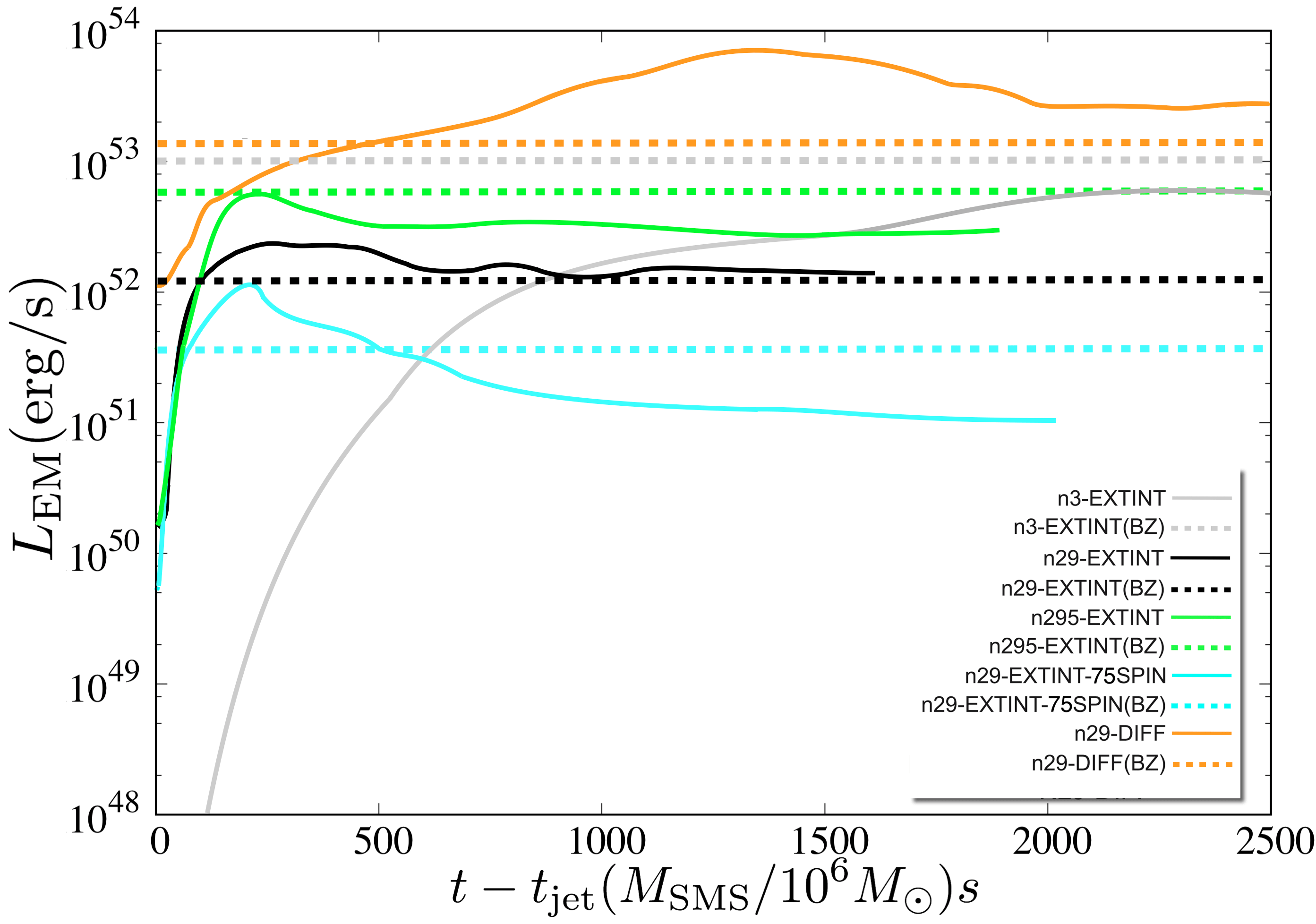}
  \caption{\label{fig:Lem} 
    Evolution of the electromagnetic Poynting luminosity $L_{\rm EM}$ crossing a sphere at coordinate
    radius $R_{ext}=100M-175M\sim 1.4-2.4\times 10^8 (M/10^6M_\odot)$km for all SMS models seeded with an
    external-interior magnetic
    field configuration (see Table~\ref{tab:table1}). Horizontal dashed lines indicate the expected BZ
    values computed via Eq.~(\ref{eq:EM_BZ}). Here $t_{jet}$ is the time at which
    the jet front has reached $\sim 100M$ above the black hole pole.}
\end{figure}

During stellar contraction and black hole formation,
the magnetic field winds up, causing the magnetic pressure to grow. A reverse shock pushes away
material that is tied to the disk via the frozen-in magnetic field
lines, producing a strong poloidal magnetic field, as shown in the right-middle panel
of Fig.~\ref{fig:Evolution_diff}. As pointed
out in~\cite{SunPasRui17,PasRuiSha15}, the conversion of poloidal to toroidal flux via magnetic
winding produces large magnetic pressure gradients above the BH that eventually launches
a strong outflow  sustained by helical magnetic fields (see the bottom panel of Fig.~\ref{fig:Evolution_diff}).
In the following we summarize additional differences in the evolution of the models
listed in table~\ref{tab:table1}.
%
%
\paragraph{\bf Models spinning at the mass-shedding limit:} Except for  n29-INT,
in which we do not observe any indication of jet formation, the early evolution and
outcome of the uniform rotating SMSs spinning at the  mass-shedding limit is
similar~(see the first three panels of Fig.~\ref{fig:initial}); at about $t - t_{\rm BH}
\approx 250-550M\sim 1.2-2.7 \times 10^3(M/10^6M_\odot)$s an incipient jet is launched following
the growth of  magnetic of pressure gradients above the black hole poles (for details see
Table~\ref{tab:table2}).  As is shown in Fig.~\ref{fig:Mdot}, the accretion rate in all these cases
settles to roughly $\sim 1\,M_\odot/s$ by about
$t-t_{BH}\sim 1.8\times 10^3M \sim 8.8\times 10^3(M/10^6M_\odot)$s, at which the mass of the disk
is $M_{disk}\sim 1.5-7.0\times 10^4(M/10^6M_\odot)M_\odot$. Hence, the duration of the jet is $\Delta
t= M_{disk}/\dot{M}\sim 1.8-7.2\times10^4(M/10^6M_\odot)$s, consistent with estimates of ultra-long
gamma-ray bursts (ULGRBs) duration in~\cite{Matsumoto:2015bga,Matsumoto:2015bjg}.
To verify that the  BZ mechanism is operating in our systems, we compare the Poynting luminosity $L_{EM}$
computed through Eq.~(7) in~\cite{SunPasRui17}  with the expected EM power generated by BZ~\cite{BlaZna77},
\begin{align}
  \label{eq:EM_BZ}
  L_{BZ}&\approx\\
  &10^{51}\left(\frac{\tiny{a/M_{BH}}}{0.75}\right)^2\,
  \left(\frac{M_{BH}}{10^6M_{\odot}}\right)^2\,
  \left(\frac{B^{\rm pole}_{BH}}{10^{10}\text{G}}\right)^2\,\rm{erg/s}\,.
  \nonumber
\end{align}
As in~\cite{SunPasRui17}, the magnetic field $B^{\rm pole}_{BH}$ is computed as a space- and  time-averaged
value of the field in a cubical region with a side of length $2\,r_{BH}$, where $r_{BH}$
is the radius of the AH, just above the black hole poles over the last $t\approx 200M\sim 1000 (M/10^6M
_{\odot})$s after the jet is well-developed. As it is displayed in Fig.~\ref{fig:Lem}, the outgoing
electromagnetic Poynting luminosity passing through a sphere with coordinate radius $R_{ext}=100M\sim 1.4\times
10^8(M/10^6M_\odot)$km is $L_{EM}\approx 10^{51}-10^{52}\rm{erg/s}$, roughly consistent with the expected BZ
value. We also compute the ratio of the angular frequency of the magnetic field lines to
the black hole angular frequency  $\Omega_F/\Omega_H$ in magnetically dominated regions above
the black hole poles (see Eq.(9) in~\cite{SunPasRui17}). We find that in these cases the ratio is
$\Omega_F/\Omega_H\approx 0.2-0.4$. Deviations from the expected split-monopole force-free magnetic
field configuration value $\Omega_F/\Omega_H=0.5$~\cite{McK20004} are expected due to differences in the
field topology and other numerical artifacts~(see e.g.~\cite{PasRuiSha15,RuiLanPas16}). The helical
structure of the polar B-field and the collimation of the outflow further suggest that the BZ
mechanism is operating in our simulations.

The lack of an incipient jet in the n29-INT model might be due to the fact that during the
stellar collapse, the black hole swallows almost the entire star, leaving only $\sim 1\%$ of the rest
mass of the SMS as a disk. During that process, the highly magnetized layers of the SMSs are captured,
leaving only the very outer layers, which are weakly magnetized, to form the remnant disk. By contrast,
the outer layer in the remnant disk in the n29-EXTINT model is highly magnetized. Following the 
collapse, we find that the magnetic field strength above the black hole poles is
$\lesssim 10^8(10^6M_\odot/M)$G in the n29-INT model case, while in the n29-EXTINT case it approaches
$\sim 10^{10} (10^6M_\odot/M)$G. As it has been pointed out in ~\cite{SunPasRui17}, the other significant difference
is  that  configurations in which the magnetic field extends from the stellar interior to its exterior mimic
a force-free environment more accurately, and as a result it is easier for the magnetic pressure to overcome
the plasma ram pressure because of less baryon loading.  Following the appearance of an apparent horizon,
we trace the plasma parameter $b^2/2\rho_0 = B^2/(8 \pi\,\rho_0)$ (where $B$ is the comoving magnetic field
strength) in a cubical region above the black hole poles. This parameter measured the degree to which the region
above the BH poles is force-free. Values larger than $\sim 1 - 10$ are required to launch a jet. We observe
that in the n29-INT model, the plasma parameter rapidly settles down to $\sim 10^{-5}$, while in the other cases
it reaches values larger than $25$ (see Table~\ref{tab:table2}). As we have seen in ~\cite{SunPasRui17}, because
of the weakly magnetized outer layer, it takes twice as long for the n3-INT cases to build up the jet than n3-EXTINT,
which might also be true for $n = 2.9$ cases. However, the computational resources required for this is overly
expensive.
%
%
%
\paragraph{\bf Model spinning at half of the mass-shedding limit:}
The evolution and final outcome of the uniformly rotating SMS spinning at half of the mass-shedding limit
is qualitatively the same as those at the mass-shedding limit (see bottom panel of Fig.~\ref{fig:initial}).
Following pressure depletion, the star shrinks and forms a central core that undergoes collapse. A black hole
horizon appears about $t_{BH} = 6.5\times 10^3M\sim 3.25\times 10^4 (M/10^6
M_\odot)$s, slightly later than in the mass-shedding limit case, because this SMS model is a less compact than
the previous cases (see Table~\ref{tab:table1}). Due to less centrifugal force,  during the first  episode
of high accretion the star is rapidly swallowed by the black hole, leaving only a tiny cloud of magnetized gas
consisting of only $\sim 0.3\%$ of the rest mass of the star  (see bottom panel of Fig.~\ref{fig:BH_m_a}). Following
the high accretion episode, the remnant magnetic field lines wind up, and around $t_{BH}\approx 340\sim 1.2\times 10^3(M/10^6
M_\odot)$s the system launches a jet. The accretion rate settles down to $1.5M_\odot/s$ and, therefore the disk is
expected to last for an accretion time $t= M_{disk}/\dot{M}\sim 1\,\times10^4(M/10^6M_\odot)$s. These numbers once again
are consistent with observations of ULGRBs~\cite{Matsumoto:2015bga,Matsumoto:2015bjg}. Once the jet is well--developed,
the time--averaged  Poynting luminosity over the last $200M\sim 10^3(M/10^6M_\odot)$s crossing a sphere at coordinate
radius $R_{ext}=100M\sim 1.4\times 10^8(M/10^6M_\odot)$km is $L_{EM}\approx 10^{51.5}\rm{erg/s}$, roughly consistent
with the expected BZ luminosity (see Fig.~\ref{fig:Lem}). At this time, the force-free parameter has reached a
value of $b^2/(2\,\rho_0) \sim 60$. As it can be seen in Eq.~(\ref{eq:EM_BZ}), a lower luminosity for this case
is expected according to Eq.~(\ref{eq:EM_BZ}) because of the  lower spin. 
%
%
\paragraph{\bf Differentially rotating model:}    
As already mentioned, differential rotation provides a centrifugal barrier to collapse. However,
redistribution of angular momentum occurs due to magnetic winding, followed by transport by
turbulent viscosity arising from MRI. Viscosity drives the external layers of the SMS outward and
the inner core toward uniform rotation (see top right panel of Fig.~\ref{fig:Evolution_diff})
allowing the inner core to collapse. At around $t_{BH} = 2.5\times 10^4M\sim 1.3
\times 10^5 (M/10^6M_{\odot})$s a black hole forms surrounded by a denser, highly spinning,
magnetized cloud of gas (see middle panels of Fig.~\ref{fig:Evolution_diff}). Unlike the  previous
$n=2.9$ models, differential rotation prevents not only the outermost layers of the star to be
accreted onto the black hole,  but also some of the inner and more magnetized layers. We find that
the average magnetic field strength at the pole is $\sim 10^{11} (10^6M_\odot/M)$G. The incipient
jet is launched at $t_{BH}\approx 286.2M\sim 1.4\times 10^3(M/10^6 M_\odot)$s, i.e. approximately
at the same time as in the previous cases (see bottom panels of Fig.~\ref{fig:Evolution_diff}).
Following the black hole formation, we observe that the plasma parameter  grows rapidly and
reaches values of $b^2/2\rho_0 \gtrsim 100$. Note that, as it has been previously discussed
in~\cite{PasRuiSha15,RuiLanPas16}, our numerical approach may be not reliable for higher  values
of the plasma parameter ($\gtrsim 200$), but the growth of magnetization in the funnel is robust,
and thus is the magnetically sustained outflow. Finally, the outgoing Poynting luminosity compute
is  $L_{\rm EM} \approx 10^{53.5} \rm erg/s$, consistent with the BZ mechanism (see Fig.
\ref{fig:Lem}).  At late times the accretion rate settles down to $\sim 2.0 M_{\odot}/s$.
The jet duration is thus $t\sim 9.0\times 10^4 (M/10^6M_\odot)$s (see table \ref{tab:table2}),
again consistent with ULGRBs observations, which may have Pop III stars as progenitors
\cite{Fermi-LAT13,TomYooBro16}.
%
\begin{table*}[!ht]
  \caption{\label{tab:model_jet} Order of magnitude comparison of simulation results with the unified model of~\cite{Sha17}.}
  \begin{center}
    \begin{tabular}{cccccccccccccccccc} 
      \hline\hline
      \multirow{2}{*}{Case}  &
      \multicolumn{2}{c}{\underline{\hspace{0.30cm} $\rho M_{BH}^2$\hspace{0.30cm}} }     &
      \multicolumn{2}{c}{\underline{\hspace{0.30cm} $B_p^2 M_{BH}^2$\hspace{0.30cm}} }    & 
      \multicolumn{2}{c}{\underline{\hspace{0.30cm} $\dot{M}_{\rm eq}(M_{\odot}/\rm s)$\hspace{0.30cm}}} &
      \multicolumn{2}{c}{\underline{\hspace{0.30cm} $\tau_{disk}/M_{BH}$\hspace{0.30cm}} }&
        \multicolumn{3}{c}{\underline{\hspace{0.30cm} $L_{EM}(\rm erg/s)$\hspace{0.30cm}} } \\
      & model~\footnote{\label{footnote1} Use Eqs.(9-12) and (17) in ~\cite{Sha17}.}  & simulations & model~\footref{footnote1} & simulations & model~\footref{footnote1}   & simulations & model~\footref{footnote1}   & simulations  & model ~\footref{footnote1}  &model BZ ~\footnote{Use Eq.~\ref{eq:EM_BZ}.} &simulations\\
      \hline\hline
	n3-HYD               & $10^{-7}$  & $10^{-7}$  & -    & -     &  $10^0$    &$10^0$   & $10^5$    & $10^5$ &  -        & -    & -         \\
	n3-INT               & $ 10^{-7}$ & $10^{-7}$  &  $10^{-6}$ & $10^{-6}$ & $10^0$    &$10^0$   & $10^{5}$  & $10^5$ &$10^{52}$ &$10^{52}$   & $10^{51}$ \\
	n3-EXTINT            & $ 10^{-8}$ & $10^{-8}$  &  $10^{-6}$ & $10^{-6}$ & $10^0$    &$10^0$   & $10^{5}$  & $10^5$ & $10^{52}$& $10^{53}$   & $10^{53}$ \\
	n295-EXTINT          & $10^{-7}$  & $10^{-6}$  & $10^{-6}$ & $10^{-6}$ & $10^0$    &$10^0$   & $10^4$    & $10^4$ & $10^{52}$ &$10^{52}$   & $10^{52}$  \\
	n29-HYD              & $ 10^{-9}$ & $10^{-9}$      &  -        & -         & $10^0$    &$10^0$   & $10^{5}$  & $10^3$ & -         &-   & -    \\
	n29-INT              & $ 10^{-8}$ & $10^{-6}$  & $10^{-6}$ & $10^{-7}$ & $10^0$    &$10^{-1}$& $10^{4}$  & $10^5$  & $10^{51}$ &$10^{51}$   & $10^{50}$\\
	n29-EXTINT           & $10^{-8}$  & $10^{-7}$ & $10^{-6}$ & $10^{-7}$ & $10^0$    &$10^0$   & $10^4$    & $10^4$  & $10^{52}$&$10^{52}$    & $10^{52}$ \\
	n29-EXTINT-0.75SPIN    & $ 10^{-8}$ & $10^{-7}$  & $10^{-8}$ & $10^{-7}$ & $10^{-1}$ &$10^0$   & $10^{3}$  & $10^2$& $10^{51}$ &$10^{51}$   & $10^{52}$ \\
	n29-EXTINT-DIFF      & $ 10^{-8}$ & $10^{-8}$   & $10^{-7}$ & $10^{-7}$ & $10^1$    &$10^0$   & $10^{6}$  & $10^6$& $10^{53}$&$10^{53}$    & $10^{54}$\\
	\hline
	\end{tabular}
  \end{center}
\end{table*}

%
%
\subsection{Comparison with the unified analytic model}
\label{sec:uni-model}
Spinning black holes immersed in magnetized accretion disks that launch collimated jets
confined by helical magnetic fields from their poles were found via our numerical
simulations to be the outcomes of three different scenarios: binary black hole-neutron
star mergers ~\cite{PasRuiSha15}, binary neutron star mergers~\cite{RuiLanPas16} and SMS
collapse~\cite{SunPasRui17}. Surprisingly, while these all represent very different scenarios
involving objects spanning a huge range of masses, length and time scale, the final quasistationary
Poynting luminosities from the jets and the mass accretion rates onto the black holes were all within
a few magnitudes of each other! This finding was recently explained by a simple analysis~\cite{Sha17}
where we showed that all the results could be understood in terms of the following universal relations:
\begin{align}
\label{eq:EM_BZ-2}
  L_{\rm BZ} &\sim \frac{1}{10}\left(\frac{M_{\rm disk}}{M_{\rm BH}}\right)
  \left(\frac{M_{\rm BH}}{R_{\rm disk}}\right)^3\left(\frac{a}{M_{\rm BH}}\right)^2 [\mathcal{L}_0]
  \nonumber\\& \sim 10^{52\pm 1}\rm erg\, s^{-1}\,
\end{align}
\begin{align}
  \dot{M}_{\rm BH} &\sim \left(\frac{M_{\rm disk}}{M_{\rm BH}}\right)\left(\frac{R_{\rm BH}}
      {R_{\rm disk}}\right)^3[\dot{\mathcal{M}}_0]
  \nonumber\\& \sim 0.1-10{M}_{\odot} \,\rm s^{-1}\,
\end{align}
where $\mathcal{L}_0\equiv c^5/G = 3.6 \times 10^{59} \rm erg \, s^{-1}$ and $\dot{\mathcal{M}}_0
\equiv c^3/G = 2.0 \times 10^5 M_{\odot} \,\rm s^{-1}$. Table~\ref{tab:model_jet} shows a comparison with
these model predictions. We find that in within one order of magnitude, the model is consistent with the
numerical results reported in this paper. Therefore, it provides another proof that the EM mechanism
running in our cases is mainly based on the BZ mechanism, on which the analysis in~\cite{Sha17} is based,
and it indicates the universality of the EM luminosity from these different scenarios. Additionally, the
EM signatures obtained from our models indicate consistency with the spectroscopic measurements from a
recent survey of short and long GRBs~\cite{LiZhaLu16}.
%
%
\subsection{GW signals and PPI in the BH-disk system}
\label{sec:PPI}
\begin{figure}[h]
\includegraphics[scale=0.12]{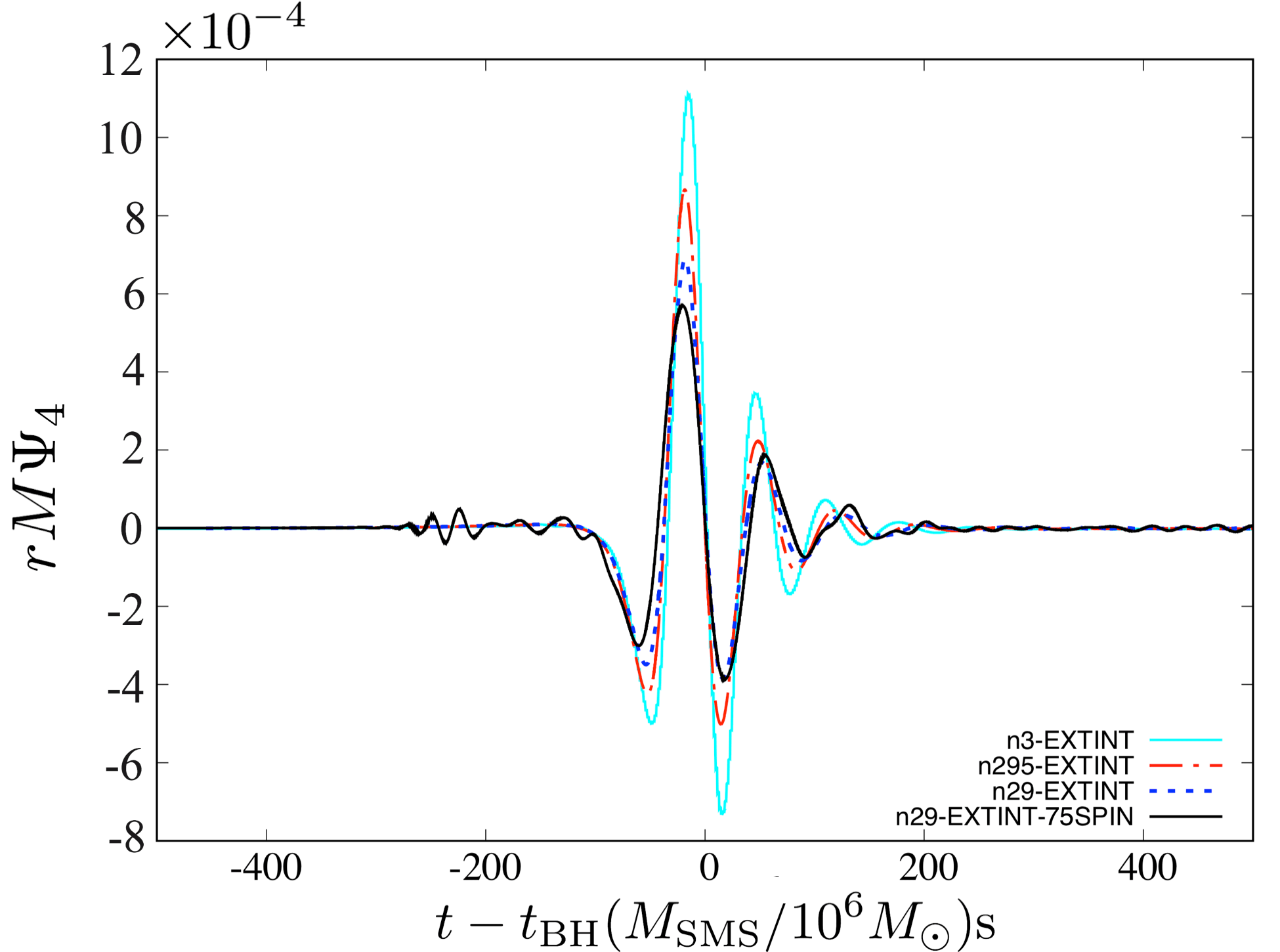}
\caption{\label{fig:Psi4} Real part of the $(l, m) = (2, 0)$ mode of $\Psi_4$ as a function of
  $t-t_{BH}$ for the cases with interior and exterior B-field at an extraction
      radius $R_{\rm ext}\sim 100M$. The cyan curve represents the
  n3-EXTINT case displayed in ~\cite{SunPasRui17}. Cases with other B-field configurations share
  similarity with their counterparts in the figure.}
\end{figure}

To extract the gravitational wave, we project the Weyl scalar $\Psi_4$
onto different extraction spheres with radii from $R_{\rm ext}\sim 100M$ to
$400M$,  and describe its angular dependence in terms of s=-2 spin-weighted spherical harmonics
(see Eqs. (3.5) and (3.6) in~\cite{RuiAlcNun07}). Fig~\ref{fig:Psi4} shows the dominant mode
$(l=2,m=0)$ of the expansion coefficient
$\Psi_4(t,r)$ at an extraction radius $R_{\rm ext}\sim 100M$  for all  cases
with interior and exterior B-fields. We find that
the peak amplitudes of $\Psi_4$ for all the cases are between $0.5-0.8$ times that of
the $n = 3$ cases, decreasing with decreasing $n$. The reason is that critical configurations
with smaller $n$ have larger compaction, hence they acquire a smaller infall speed at collapse.
The oscillation period of this mode in all cases resembles the $n = 3$ waveform $(f \sim 15
(10^6M_{\odot}/M)/(1+z) \rm mHz)$ and both amplitude and frequency are consistent with the results
obtained from the axisymmetric SMS collapse reported in~\cite{ShiSekUch16}.
Therefore, analogous to the discussion of detectability for the $\Gamma = 4/3$ cases in
\cite{SunPasRui17}, it is expected that GW detectors most sensitive to the $10^{-3} -
10^{-1} \rm Hz$ band (e.g. LISA and DECIGO) are able to observe the GW signals from such systems
\cite{eLISA_sensitivity, Yagi:2009zz,Yagi:2011yu}. 
It was suggested by~\cite{KiuShiMon11} that a detectable, quasiperiodic post-collapse signal
might arise from the BH-disk system due to the growth of the $m=1$ nonaxisymmetric
Papaloizou-Pringle instability (PPI). However, we find that compared to the $(l,m) = (2,0)$
mode, all nonaxisymmetric modes are significantly smaller. This suggests that no pronounced
and sustained oscillatory waveform is produced in this system, in contrast to Fig. (3) in
\cite{KiuShiMon11}. Therefore, PPI and its associated GW signal do not arise in our simulations.
The reason that the instability is absent could be either that the BH-disk system is stable with
respect to PPI even in the absence of a magnetic field, or that the instability is suppressed by
the magnetic fields and the development of MRI ~\cite{BugGuiMul17}. To address this question we
compute the specific angular momentum profile $j = u^t u_\phi$ versus $r$ in the equatorial plane.
If $\Omega \sim  r^{-q}$, then in the Newtonian limit,
\begin{equation}
j \sim \Omega r^2 \sim r^{2-q},
\end{equation}         
in which case we know that the disk in the absence of magnetic fields is unstable whenever
$q > \sqrt{3}$ or $2-q < 0.27$~\cite{PapPri85}. We find that for nonmagnetized cases, our post-collapse
quasiequilibrium disks satisfy $ 2-q \sim 0.33-0.35$, which suggests that the disks are stable, even
in the absence of magnetic fields. Furthermore, disks formed by collapsing magnetized models result
in $ 2-q \sim 0.47-0.55$, in which case the disk stability with respect to PPI may be enhanced by the
magnetic field~\cite{BugGuiMul17}.
%
\section{Summary and Conclusions}
\label{sec:Summ-Conc}
In this work, we extended our earlier calculations~\cite{SunPasRui17} of the magnetorotational collapse
of SMSs in which radiation pressure alone is present($\Gamma=4/3$, or $n = 3$). Such a model applies to
SMSs with $M \gtrsim 10^6 M_{\odot}$. Hence we have performed full GRMHD simulations of collapsing of SMSs
with masses $\gtrsim 10^4 - 10^5 M_\odot$ for which gas pressure represents a significant perturbation
\cite{BonArnCar84,ShiSekUch16,Uchida:2017qwn}. We considered stellar models described by
a  polytropic EOS with $\Gamma \gtrsim 4/3$, or equivalently $n \lesssim 3$, which effectively incorporates
a gas pressure perturbation. Such a model also crudely describes massive Pop III stars. We also studied the
impact of the initial stellar rotation profile and the initial magnetic field
configuration on the final outcome of the SMS remnant. To be consistent with our previous study
\cite{SunPasRui17}, we set the initial magnetic-to-rotational-kinetic energy to $0.1$. 

We focus on uniformly rotating configurations spinning at mass-shedding and on the verge of collapse
due to a relativistic radial instability. For uniformly rotating cases, the
evolution process is similar to the $n = 3$ cases presented in~\cite{SunPasRui17} with the same initial
magnetic field configuration. For smaller characteristic masses (smaller $n$), the stars collapse
in a shorter period. The outcome in all cases is a spinning black hole surrounded by an accretion disk.
For smaller initial $n$ and thus smaller $M$, the BH has a greater $M_{\rm BH}/M$ and thus a smaller
$M_{\rm disk}/M$, and a smaller $a_{\rm BH}/M_{\rm BH}$. All the black hole parameters are consistent with
various previous semi-analytic and numerical studies in~\cite{CooShaTeu94, BauSha99,Sha04,ShiSha02,LiuShaSte07}.
For SMSs with $M \gtrsim 10^6 M_{\odot}$, the ratios $M_{\rm BH}/M$, $a_{\rm BH}/M_{\rm BH}$ and $M_{\rm disk}/M$ are
universal numbers independent of mass ~\cite{ShaShi02}: $M_{\rm BH}/M \approx 0.9$,  $a_{\rm BH}/M_{\rm BH}
\approx 0.75$ and $M_{\rm disk}/M \approx 0.1$. Furthermore, for all magnetized cases, the final
$\dot{M}_{\rm BH}$ is roughly the same ($\sim ~ 0.1-1 M_{\odot}/\rm s$) as is the Poynting luminosity
($L_{\rm EM}\sim 10^{52 \pm 1} \rm erg/s$), independent of $M$. These are consistent with the $n = 3$ cases and
with the unified analytic model in~\cite{Sha17}.

For the cases with reduced spin $\Omega = 0.75 \Omega_{\rm shedd}$, we found that almost all the matter
falls into the black hole, with only $\sim 0.3\%$ of the total mass remaining to form the disk. Correspondingly, $L_{\rm EM}$
is approximately one order of magnitude smaller than its uniformly rotating counterpart. On the contrary,
the collapse of a differentially rotating star results in a massive disk with $M_{\rm disk}/M \sim 0.18$ and
the highest luminosity $L_{\rm EM} \sim 10^{53.5} \rm erg/s$. 

We find that all appreciably magnetized disks launch incipient jets. We confirm the likelihood that the BZ mechanism
generates the Poynting luminosity in the jets. The gravitational waveforms for $n \lesssim 3$ show strong resemblance
to their $n = 3$ counterparts. It is thus expected that GW detectors like LISA and DECIGO are capable of observing the GW
signals from such events~\cite{SunPasRui17}. The specific angular momentum profiles in the post-BH disk show that the disk
is stable with respect to PPI even without the magnetic field, and such stability is probably
strengthened in presence of the magnetic field. Additionally, the magnitude of the Poynting luminosity, which is insensitive
to the stellar mass $M$, suggests that detecting the EM counterpart radiation from magnetized, massive, stellar collapses
by GRB detectors like \textit{Fermi} and \textit{Swift} is quite feasible~\cite{BosGotBou14, ConPelBri13}. Therefore, the
study of and search for SMSs or massive Pop III stars could provide a promising avenue for advancing multimessenger
astronomy research.

An extensive survey of different rotation profiles is clearly needed to strengthen our conclusion
(e.g. Eq.~\ref{eq:EM_BZ-2}) regarding the EM luminosity in the case of differentially rotating stars.
However, the results reported here,  along  with the simulations of supermassive black holes
surrounded by accretion disk in~\cite{Khan:2018ejm}, the simulations of black hole-neutron star mergers
in~\cite{PasRuiSha15}, and those of binary neutron star mergers~\cite{RuiLanPas16} suggest that indeed
there maybe a narrow range of expected EM luminosity in accord with Eq.~\ref{eq:EM_BZ-2} and the
analysis in~\cite{Sha17}.
\acknowledgements
We thank V. Paschalidis for useful discussions, and the Illinois Relativity Group
REU team (E. Connelly, J. Simone and I. Sultan) for visualization assistance. This work has been
supported in part by National Science Foundation (NSF) Grants
PHY-1602536 and PHY-1662211, and NASA Grant 80NSSC17K0070 at the University of Illinois
at Urbana-Champaign. This work made use of the Extreme Science and Engineering Discovery Environment (XSEDE),
which is supported by National Science Foundation grant number TG-MCA99S008. This research is part of the Blue
Waters sustained-petascale computing project, which is supported by the National Science Foundation (awards
OCI-0725070 and ACI-1238993) and the State of Illinois. Blue Waters is a joint effort of the University of
Illinois at Urbana-Champaign and its National Center for Supercomputing Applications.

\bibliographystyle{apsrev}
\bibliography{references}

\end{document}